\documentclass[english]{article}
\pdfoutput=1
\usepackage[T1]{fontenc}
\usepackage{geometry}
\usepackage{bm}
\usepackage{amsmath}
\usepackage{amssymb}
\usepackage{esint}
\usepackage{graphicx}
\usepackage{amsthm}

\usepackage{color}
\definecolor{darkblue}{rgb}{0,0,0.6}
\definecolor{darkred}{rgb}{0.6,0,0}
\definecolor{darkgreen}{rgb}{0,0.6,0}
\usepackage[colorlinks=true,urlcolor=darkblue,citecolor=darkblue,linkcolor=darkred,hyperfootnotes=false]{hyperref}

\makeatletter
\usepackage{cite}
\usepackage[english]{babel}
\usepackage[T1]{fontenc}
\usepackage[latin1]{inputenc}
\IfFileExists{lmodern.sty}{\usepackage{lmodern}}{}
\usepackage{amsmath}
\usepackage{braket}
\usepackage{amssymb}
\usepackage{dsfont}
\usepackage{setspace}
\usepackage{authblk}
\usepackage{lipsum}
\RequirePackage{doi}
\usepackage{graphicx}
\usepackage{bibspacing}
\usepackage{url}
\makeatother

\newtheorem*{theorem*}{Rule}
\newcommand\blfootnote[1]{%
  \begingroup
  \renewcommand\thefootnote{}\footnote{#1}%
  \addtocounter{footnote}{-1}%
  \endgroup
}

\usepackage{babel}

\begin{document}
\global\long\def\dbar{\mathchar'26\mkern-12mud}

\global\long\def\bnabla{\bm{\nabla}}

\global\long\def\lap{\nabla^{2}}

\global\long\def\r{\mathbb{R}}

\global\long\def\FancyH{\mathcal{H}}

\global\long\def\FancyF{\mathcal{F}}

\global\long\def\FancyG{\mathcal{G}}

\global\long\def\FancyD{\mathcal{D}}

\global\long\def\angs{\mathring{\mathrm{A}}}

\global\long\def\c{\mathbb{C}}

\global\long\def\n{\mathbb{N}}

\global\long\def\z{\mathbb{Z}}

\global\long\def\q{\mathbb{Q}}

\global\long\def\f{\mathbb{F}}

\global\long\def\id{\mathds{1}}

\global\long\def\tr{\mathrm{Tr}\thinspace}

\global\long\def\det{\mathrm{det}}

\global\long\def\dim{\mathrm{dim}}

\global\long\def\ker{\mathrm{ker}}

\global\long\def\im{\mathrm{Im}}

\global\long\def\hom{\mathrm{hom}}

\global\long\def\mo#1{\mathcal{O}\!\left(#1\right)}

\global\long\def\rlim#1#2#3{\xrightarrow[#1\rightarrow#2]{#3}}

\global\long\def\clim#1#2{\lim_{#1\rightarrow#2}}

\global\long\def\llim#1#2#3{\begin{array}{c}
 #3\\
 \ll\\
 #1\rightarrow#2 
\end{array}}

\global\long\def\glim#1#2#3{\begin{array}{c}
 #3\\
 \gg\\
 #1\rightarrow#2 
\end{array}}

\global\long\def\aseq#1#2#3{\begin{array}{c}
 #3\\
 \sim\\
 #1\rightarrow#2 
\end{array}}

\global\long\def\grandz{\mathcal{Z}}

\global\long\def\const{\mathrm{const}}

\global\long\def\bp{\mathbf{p}}

\global\long\def\bv{\mathbf{v}}

\global\long\def\bx{\mathbf{x}}

\global\long\def\bk{\mathbf{k}}

\global\long\def\bq{\mathbf{q}}

\global\long\def\br{\mathbf{r}}

\global\long\def\mb{\mathbf{m}}

\global\long\def\bn{\mathbf{n}}

\global\long\def\Mb{\mathbf{M}}

\global\long\def\bl{\mathbf{L}}

\global\long\def\bj{\mathbf{J}}

\global\long\def\div{\bnabla\cdot}

\global\long\def\rot{\bnabla\times}

\global\long\def\braket#1{\left\langle #1\right\rangle }

\global\long\def\col#1{\left\langle #1\right\rangle _{c}}

\global\long\def\ket#1{\left|#1\right\rangle }

\global\long\def\bra#1{\left\langle #1\right|}

\global\long\def\abs#1{\left|#1\right|}

\global\long\def\absinf#1{\left|#1\right|_{\infty}}

\global\long\def\abstwo#1{\left|#1\right|_{2}}

\global\long\def\absone#1{\left|#1\right|_{1}}

\global\long\def\absp#1#2{\left|#1\right|_{#2}}

\global\long\def\norm#1{\left\Vert #1\right\Vert }

\global\long\def\bracket#1{\left[#1\right]}

\global\long\def\curly#1{\left\{  #1\right\}  }

\global\long\def\subs#1{\left.#1\right|}

\global\long\def\vint#1{\int d^{#1}\mathbf{r}}

\global\long\def\sint{\int d\mathbf{S}\cdot}

\global\long\def\lint{\int d\mathbf{l}\cdot}

\global\long\def\clint{\oint d\mathbf{l}\cdot}

\title{Bodies in an Interacting Active Fluid: Far-Field Influence\\of a Single Body and Interaction Between Two Bodies}
\author[1,*]{Omer Granek}
\author[2]{Yongjoo Baek}
\author[1]{Yariv Kafri}
\author[3]{Alexandre P. Solon}
\affil[1]{\small{\textit{\normalsize{Department of Physics, Technion - Israel Institute of Technology, Haifa 32000, Israel}}}}
\affil[2]{\small{\textit{\normalsize{Department of Physics and Astronomy, Seoul National University, Seoul 08826, Korea}}}}
\affil[3]{\small{\textit{\normalsize{Sorbonne Universit{\'e}, CNRS, Laboratoire de Physique Th{\'e}orique de la Mati{\`e}re Condens{\'e}e, LPTMC, F-75005 Paris, France}}}}
\maketitle

\begin{abstract}
\blfootnote{\textsuperscript{*} \href{mailto:omer.granek@campus.technion.ac.il}{omer.granek@campus.technion.ac.il}}Because active particles break time-reversal symmetry, an active fluid
can sustain currents even without an external drive. We show that when a
passive body is placed in a fluid of pairwise interacting active particles,
it generates long-range currents, corresponding to density and pressure
gradients. By using a multipole expansion and a far-field constitutive
relation, we show that the leading-order behavior of all three corresponds
to a source dipole. Then, when two bodies or more are placed in the
active fluid, generic long-range interactions between the bodies occur.
We find these to be qualitatively different from other fluid mediated
interactions, such as hydrodynamic or thermal Casimir. The interactions
can be predicted by measuring a few single-body properties in separate
experiments. Moreover, they are anisotropic and do not satisfy an
action-reaction principle. These results extend previous results on non-interacting active particles. Our framework may point to a path towards self-assembly.
\end{abstract}

\tableofcontents

\section{Introduction}

{\it Active particles}, which propel themselves by consuming stored or ambient energy, form an interesting class of far-from-equilibrium systems~\cite{RamaswamyARCMP2010,MarchettiRMP2013,BechingerRMP2016,RamaswamyJSM2017}. They have attracted much attention due to unusual collective phenomena which are not found in equilibrium. Examples include flocking~\cite{VicsekPhysRep2012,DeseignePRL2010,DeseigneSM2012,solon2013revisiting,toner1998flocks,toner2005hydrodynamics}, motility-induced phase separation (MIPS)~\cite{fily2012athermal,CatesARCMP2015,SolonPRL2015,SolonPRE2018,SolonNJP2018,DigregorioPRL2018}, and the lack of an equation of state for pressure~\cite{SolonNatPhys2015,NikolaPRL2016,JunotPRL2017}. In particular, it is known that asymmetric obstacles immersed in a fluid of active particles (called an {\it active fluid}) create density gradients~\cite{GalajdaJBact2007,GalajdaJMO2008,GuidobaldiPRE2014} and currents~\cite{ReichhardtPRL2008,MahmudNatPhys2009,KantslerPNAS2013,GuidobaldiPRE2014}. These phenomena are examples of the {\it ratchet effect}: directed motion can be extracted out of fluctuations by breaking both spatial symmetry and time-reversal symmetry \cite{MagnascoPRL1993,ProstPRL1994}. The breaking of spatial symmetry is provided by the asymmetric obstacle, while the breaking of time-reversal symmetry stems from the dynamics of the active particles \cite{TailleurEPL2009,BechingerRMP2016,ReichhardtARCMP2017}. The active ratchet mechanism can be applied to targeted delivery~\cite{KoumakisNatComms2013}, self-starting micromotors~\cite{AngelaniPRL2009,DiLeonardoPNAS2010,KaiserPRL2014} and self-assembly of colloidal molecules~\cite{SotoPRL2014,SotoPRE2015,MalloryARPC2018}.

Recently, the effects of {\it arbitrary} asymmetric bodies immersed in a non-interacting fluid of Run-and-Tumble or Active Brownian Particles were analyzed and quantified~\cite{BaekPRL2018}. First, it was shown that even a single localized asymmetric body generates a long-range density disturbance which decays as a power law and whose structure is mathematically similar to the potential of an electric source dipole. The strength of the dipole is directly related to the force exerted by the body on the active fluid. In turn, a current field, whose far-field behavior is similar to the field of an electric source dipole, is generated. We note that a similar mechanism was also found to exist in diffusive systems with an asymmetric localized drive~\cite{BodineauJSP2010,SadhuPRE2011,SadhuPRE2014,SadhuJPA2014,CividiniPRE2017}.

This led to the finding that, when multiple bodies are placed in the fluid, long-range interactions exist between the bodies. These interactions, expressed through forces and torques, are long range with a magnitude decaying with distance as a power law. They are directly related to the density and current fields produced by a single body. Hence, they differ from the previously observed confinement-induced interactions~\cite{AngelaniPRL2011,RayPRE2014,HarderJCP2014,RanPRL2015,LeitePRE2016}, which decay over a finite characteristic length-scale. The interactions are different from the conventional long-range interactions, such as hydrodynamic interactions~\cite{HappelBrenner1983} and similar bath-mediated interactions~\cite{ReichhardtPRE2006,MejiaSM2011,VissersSM2011,LadadwaPRE2013,VasileyvSM2017,PoncetARXIV2019}, which exist only among moving bodies and require interactions between the fluid particles. They also differ from thermal Casimir forces~\cite{FisherCRA1978,KardarRMP1999}, which require long-range correlations. The interactions generically exist even between static bodies in a fluid far from any critical point. The leading-order interactions were found to be fully determined by the single-body properties of each body involved, so that one can predict the interaction between a pair of bodies from separate observations of the individual isolated bodies. Moreover, the interactions fail to satisfy an action-reaction principle, showing that the activity of the particles compensates for the residual forces and torques. Such a non-Newtonian nature also exists in non-equilibrium depletion forces \cite{DzubiellaPRL2003,HayashiJPCM2006,KhairPRSA2007,PinheiroPS2011,IvlevPRX2015}. However, these are not truly scale-free, as they are screened on scales much larger than the body size \cite{KliushnychenkoPRE2017,KliushnychenkoPRE2018}. Interactions with similar scaling to this may be present in the strongly-interacting clustered phase of active matter, which exhibits almost-scale-free correlations \cite{CavagnaPNAS2010,XiaoPRL2012,WysockiEPL2014}. On the contrary, true scale-free interactions were observed following a quench in temperature in both passive and active matter \cite{RohwerPRE2018}. However, in that case, the effect is only transient. Lastly, it was demonstrated~\cite{BaekPRL2018} that the generic active fluid interactions give rise to novel dynamical phenomena involving two objects immersed in an active fluid.

In this paper, we generalize these results to the technically more demanding problem of a fluid of Run-and-Tumble or Active Brownian Particles with pairwise interactions between the active particles. Assuming that the active fluid is in a disordered phase, we show that the mathematical structure remains similar to the non-interacting problem, but with interesting differences. In particular, besides the density and current fields, one now needs to consider the pressure field. This is found to decay in a way similar to the density field, {\it i.e.}, like the potential of an electric dipole. The density field also exhibits a similar decay, but with an amplitude modified by the compressibility of the active fluid. Using the single body results, we then derive the interactions between two bodies along the lines of Ref.~\cite{BaekPRL2018}.

The paper is organized as follows. After defining the model of active particles in Sec.~\ref{sec:model}, we give a brief summary of the main results in Sec.~\ref{sec:summary}. Then, we present the derivation. First, the steady-state conditions for the active particles are shown in Sec.~\ref{sec:hydrodynamics}. These are used to obtain the far-field effects of a single body in Sec.~\ref{sec:far-field}, which in turn allows us to derive the long-range interactions between pairs of bodies in Sec.~\ref{sec:interactions}. Finally, we summarize our results and conclude in Sec.~\ref{sec:conclusions}.

\section{Model}
\label{sec:model}

We consider a model of active particles which encompasses both Active Brownian Particles (ABPs)~\cite{SchweitzerPRL1998,RomanczukEPJST2012} and Run-and-Tumble Particles (RTPs)~\cite{SchnitzerPRE1993}. The particles propel themselves at speed $v$ and interact via pairwise central forces derived from a potential $U\!\left(\abs{\br}\right)$. In what follows,
we consider only the two-dimensional case, and the generalization to
higher dimensions is straightforward. In the overdamped limit, the
position $\mathbf{r}_{i}$ and the orientation $\theta_{i}$ of active
particle $i$ are governed by the It\^o-Langevin dynamics 
\begin{align}
\dot{\mathbf{r}}_{i} & =v\mathbf{e}_{\theta_{i}}-\mu\bm{\nabla}_{\mathbf{r}_{i}}\left[V\!\left(\mathbf{r}_{i}\right)+\sum_{k\neq i}U\!\left(|\mathbf{r}_{i}-\mathbf{r}_{k}|\right)\right]+\sqrt{2D_\text{t}}\,\bm{\mathbf{\eta}}_{i}\!\left(t\right),\label{eq:langevinr}\\
\dot{\theta}_{i} & =\sqrt{2D_\text{r}}\,\xi_{i}\!\left(t\right).\label{eq:langevint}
\end{align}
Here $\mu$ is the mobility of particle $i$, $D_\text{t}$ and $D_\text{r}$ are translational
and rotational diffusion constants, the components of $\bm{\eta}_{i}$ and $\xi_{i}$
are mutually independent Gaussian white noises with unit variance, and $\mathbf{e}_{\theta_{i}}\equiv(\cos\theta_{i},\sin\theta_{i})^{T}$
is a unit vector indicating the orientation of the particle. The external potential $V$, which can be written as $V = \sum_j V_j$ with the body index $j$ in the presence of multiple bodies, describes the interaction between each active particle and the bodies immersed in the active fluid. In addition to the diffusive dynamics described by the above equations, we also allow for tumbling dynamics, {\it i.e.}, $\theta_{i}$ randomly changes its value at a rate $\alpha$. Pure ABPs correspond to $\alpha = 0$, and pure RTPs correspond to $D_\text{r} = 0$. Using this generalized model provides a unified view of active particles. In a steady state, the effect of tumbling becomes identical to the effect of active diffusion -- a property used extensively in the diffusive approximation of RTP dynamics at long time-scales \cite{SchnitzerPRE1993,OthmerSIAM2000,OthmerSIAM2002,ErbanSIAM2004,TailleurPRL2008}. This emphasizes that our results below are independent of the statistical details of the active force $v\mathbf{e}_{\theta_{i}}/\mu$. Rather, they rely on the existence of a typical distance traveled by the particle while keeping its orientation $l_\text{r}= v/\left(\alpha+D_\text{r}\right)$ (also called the {\it run length}). It is important to note that the model represents {\it dry active matter}, which is ``dry'' in the sense that it does not conserve the momentum~\cite{RamaswamyARCMP2010,MarchettiRMP2013,BenDorARXIV2018}.
Accordingly, the model best describes particles next to a surface which can absorb the momentum, such as a layer of vibrated granular particles~\cite{DeseignePRL2010,DeseigneSM2012,JunotPRL2017}
and gliding bacteria~\cite{PeruaniPRL2012}. Nonetheless, it has been shown that for this model, due to the reasons elucidated in~\cite{FilyJPA2018}, there is an equation of state for the pressure~\cite{SolonPRL2015}. This will play a salient role in the derivations that follow. 

All the results are valid in an adiabatic limit where it is assumed that the object or objects move
on a time scale much longer than the diffusive relaxation
time of the surrounding active fluid. 

\section{Main results}
\label{sec:summary}

We first review our main results before presenting their derivations in detail.
 To do so, we first consider
the case where only a single passive body is immersed in an active fluid, presenting far-field
expressions for the steady-state particle density, current density
and hydrostatic pressure field created by the body. Then we present results for the case where two passive bodies are placed at a large distance from each other in the same active fluid, giving expressions
for the forces and torques between the bodies which are mediated by the fluid.
Importantly, the interactions are expressed in terms of single-body
properties.

\subsection{Far-field effects of a single body}

We denote by $\hat{\rho}\!\left(\mathbf{r}\right)\equiv\sum_{i}\delta\!\left(\mathbf{r}-\mathbf{r}_{i}\right)$ the empirical density and by $\hat{\mb}\!\left(\br\right)\equiv\sum_{i}\delta\!\left(\mathbf{r}-\mathbf{r}_{i}\right)\mathbf{e}_{\theta_{i}}$ the empirical polarization density. A hat above a symbol indicates that the symbol stands for a random variable. The hat shall be removed after taking an average over histories, so that $y = \braket{\hat{y}}$. We use $d$ to denote the size of the body corresponding to the potential $V\!\left(\mathbf{r}\right)$. If the body is placed upon the origin of the coordinates, the far-field limit is defined as $r\gg \max\left(l_{\text{r}},d\right)$. In this limit we obtain the pressure, density, and current fields. The results are derived  assuming that (i) the active fluid is homogeneous and disordered far away from the body and (ii) the dominating component of the far-field fluid stress can be expressed as a local function of the density. We justify the second assumption in the case where, in the far field, either inter-particle interactions are weak or some correlations have a mean-field structure. Importantly, we confirm this assumption using numerical simulations which verify the theoretically predicted long-range current and density profiles. 

Denoting the modulated pressure field by $P(\mathbf{r})$, we find that it satisfies
\begin{align}
P\!\left(\br\right) & = P\!\left(\rho_{b}\right)+\frac{1}{2\pi}\frac{\mathbf{r}\cdot\mathbf{p}}{r^{2}}+\mo{r^{-2}}.\label{eq:multip}
\end{align}
Here $\rho_{b}$ is the density of active particles at $r\rightarrow\infty$, and $P\!\left(\rho_{b}\right)$ is the corresponding pressure. Throughout the paper, the decay of remainders is given up to some sub-algebraic modulation. The equation of state $P\!\left(\rho_{b}\right)$ for the pressure has been derived in a few different ways~\cite{SolonPRL2015,FilyJPA2018,SolonNJP2018,YangSM2014,TakatoriPRL2014,FalascoNJP2016} and takes the form
\begin{align}
P\!\left(\rho_{b}\right) = T_\text{eff}\,\rho_{b} + P_\text{D}\!\left(\rho_{b}\right)  + P_\text{I}\!\left(\rho_{b}\right). \label{eq:Pdec}
\end{align}
Here $T_\text{eff}\,\rho_{b}$ is the ideal-gas contribution with an effective temperature $T_\text{eff}\equiv D_\text{t}/\mu+vl_\text{r}/(2\mu)$, and $P_\text{D}$ and $P_\text{I}$ are direct and indirect contributions from the interaction potential $U$, respectively. The latter two are related to the empirical density and polarization by 
\begin{align}
P_\text{D}\!\left(\rho_{b}\right) & =-\frac{1}{4}\lim_{r\rightarrow\infty}\int d^{2}\mathbf{r}'\,r'U'\!\left(r'\right)\int_{0}^{1}d\lambda\braket{\hat{\rho}\!\left(\br+\left(1-\lambda\right)\br'\right)\hat{\rho}\!\left(\br-\lambda\br'\right)},\label{eq:pd}\\
P_\text{I}\!\left(\rho_{b}\right) & =-\frac{l_\text{r}}{2}\lim_{r\rightarrow\infty}\int d^{2}\mathbf{r}'\braket{\hat{\mathbf{m}}\!\left(\br\right)\hat{\rho}\!\left(\mathbf{r}'\right)}\cdot\bnabla U\!\left(|\mathbf{r}-\mathbf{r}'|\right),\label{eq:pi}
\end{align}
which can be written as functions of $\rho_b$ in a homogeneous and disordered fluid. We note that $P_\text{I}$ is sometimes referred to as the swim pressure of the active fluid~\cite{YangSM2014,TakatoriPRL2014}. Finally, $\mathbf{p}$ is the dipole moment given by
\begin{align}
\mathbf{p} & =-\int d^{2}\mathbf{r}\,\rho\!\left(\mathbf{r}\right)\bm{\nabla}V\!\left(\mathbf{r}\right).\label{eq:dipole-1-1}
\end{align}
It is equal to the net force applied on the fluid by the body, which is opposite and equal to the force applied on the body by the fluid. We note that $\bp=0$ for an apolar $V$, such as one with a disk-like or rod-like shape---dipole-like long-range effects are generated only if the body has a polar asymmetric shape\footnote{We note that a rod still generates long-range density modulations and currents despite having ${\bf p}=0$, see for example \cite{BaekPRL2018}. These, however, appear at quadrupolar order and are beyond the treatment carried out in the paper.}~\cite{TailleurEPL2009,NikolaPRL2016}.

Based on the above results, we also show that the average particle density $\rho\!\left(\br\right)$ can be expanded as
\begin{align} \label{eq:ffrho}
\rho\!\left(\br\right) = \rho_{b}+\frac{1}{2\pi P'\!\left(\rho_{b}\right)}\frac{\mathbf{r}\cdot\mathbf{p}}{r^{2}}+\mo{r^{-2}} = \rho_{b}\left[1 + \frac{c(\rho_{b})}{2\pi}\frac{\mathbf{r}\cdot\mathbf{p}}{r^{2}}\right]+\mo{r^{-2}},
\end{align}
where the second equality is obtained by noting that $P'(\rho_b)$ is related to the compressibility of the active fluid by the relation $c(\rho_b) = 1/\left[\rho_b \,P'(\rho_b)\right]$. In other words, for a given force (or dipole moment $\mathbf{p}$) exerted by the body on the surrounding active fluid, an active fluid of greater compressibility has greater density modulations.

Finally, the force generates a long-range current field $\bj$ whose far-field expression is given by 
\begin{align}
\bj\!\left(\br\right) & =-\frac{\mu}{2\pi}\left[\frac{\mathbf{p}}{r^{2}}-\frac{2\left(\mathbf{r}\cdot\mathbf{p}\right)\mathbf{r}}{r^{4}}\right]+\mathcal{O}\left(r^{-3}\right).\label{eq:Jff}
\end{align}

\subsection{Long-range interactions between bodies}

Building on the above results, we derive the interactions between
two passive bodies in an active fluid. We consider the case where body $2$ is separated from
body $1$ by a mutual far-field displacement $\br_{12}$. When the
system is phase separated, we assume that the two bodies are immersed
deep inside the same phase. We find that ${\bf F}_{12}$, the additional force
exerted on body $2$ due to the introduction of body $1$ into the
fluid, can be expressed by single-body properties. Specifically, we
decompose ${\bf F}_{12}$ as 
\begin{align}
{\bf F}_{12} & ={\bf F}_{12}^{a}+{\bf F}_{12}^{s},\label{eq:dec}
\end{align}
where ${\bf F}_{12}^{a}$ acts only on asymmetric bodies with non-zero
dipole moment (see Eq.~\eqref{eq:dipole-1-1}), while ${\bf F}_{12}^{s}$
is present even for fully symmetric bodies with zero dipole moment.
In what follows, we use subscript $j$ to denote the quantities appearing in the single-body problem of body $j$, {\it e.g.}, the pressure field $P_{j}\!\left(\br\right)$ and current density $\bj_{j}\!\left(\br\right)$.

\begin{figure}
  \centering
  \includegraphics[width=0.53\textwidth]{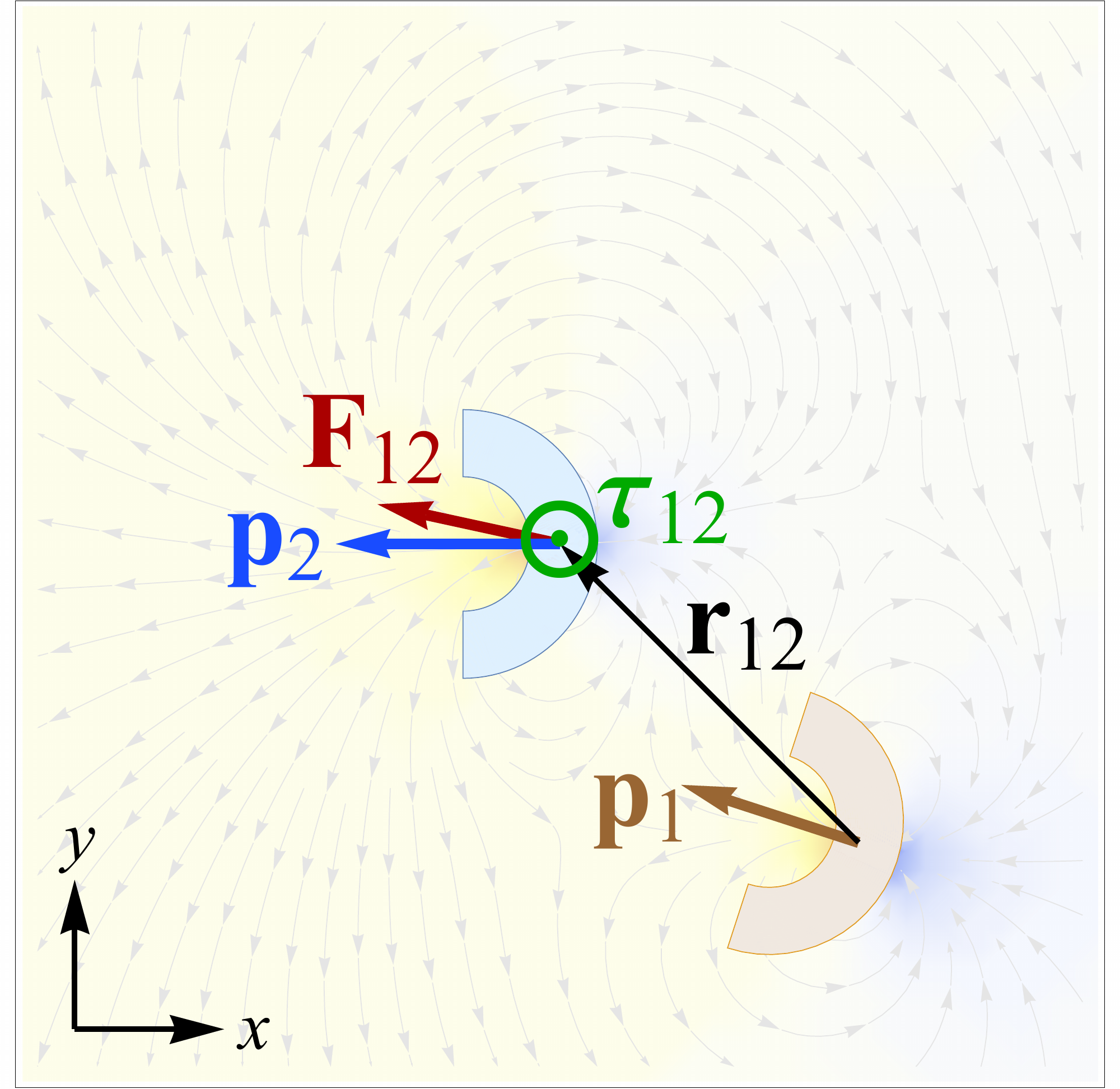}
  \caption{Schematic diagram of two interacting asymmetric passive bodies. Body $2$ (blue) is placed at the origin, while body $1$ (orange) is displaced by $\br_{12}$ (black). Superposed dipolar currents are shown in gray streamlines, and dipole density modulations are shown in red and blue colors. In this case, the linear response operators $\mathbf{R}_2^{P}$, $\mathbf{R}_2^{\rho}$, $\mathbf{T}_2$ and $\bm{\gamma}_2$ are parallel to axis spanned by the unmodified dipole moment $\bp_2$ (blue), which is the the $x$ axis. The $x$ and $y$ axes are the two principal axes of the operator $\mathbb{R}_2$. The force $\mathbf{F}_{12}$ (red) and torque $\bm{\tau}_{12}$ (green) applied on body $2$ by body $1$ are obtained from the above quantities and the unmodified dipole moment $\bp_1$. See text for results and derivations.}
  \label{fig:interaction}
\end{figure}

At leading order in $r_{12}$, ${\bf F}_{12}^{a}$ and ${\bf F}_{12}^{s}$ can be understood as the response forces induced by the pressure perturbation $\Delta P_{j}\left(\br\right) \equiv P_{j}\!\left(\br\right)-P\!\left(\rho_{b}\right)$ and the current around the body. In other words, we can write (see Fig.~\ref{fig:interaction})
\begin{align}
\mathbf{F}_{12}^{a} & =\mathbf{R}_{2}^{P}\,\Delta P_{1}\!\left(\br_{12}\right)+\mathcal{O}\!\left(r_{12}^{-2}\right),\label{eq:Fa-1}\\
\mathbf{F}_{12}^{s} & =\mathbb{R}_{2}\,\bj_{1}\!\left(\br_{12}\right)+\mathcal{O}\!\left(r_{12}^{-3}\right),\label{eq:Fs-1}
\end{align}
where we used the linear response operators defined as
\begin{align}
\mathbf{R}_{j}^{P} & \equiv\left.-\partial_{P\!\left(\rho_{b}\right)}\,\mathbf{p}_{j}\!\left[P\!\left(\rho_{b}\right),\bj_{b}\right]\right|_{\bj_{b}=0}=\int d^{2}\br\left.\partial_{P\!\left(\rho_{b}\right)}\,\rho\!\left[P\!\left(\rho_{b}\right),\bj_{b}\right]\right|_{\bj_{b}=0}\bnabla V_{j},\label{eq:response}\\
\mathbb{R}_{j} & \equiv\int d^{2}\br \left(\bnabla V_{j}\right)\left.\bnabla_{\bj_{b}}\,\rho\!\left[P\!\left(\rho_{b}\right),\bj_{b}\right]\right|_{\bj_{b}=0}.\label{eq:responseJ}
\end{align}
The response operators can be measured by
placing body $j$ alone in an active fluid, and measuring the response
of the force $-\bp_{j}$ to modified boundary conditions. This
includes modulation of the pressure $P\!\left(\rho_{b}\right)$ or application of a \textit{boundary-driven
current} $\bj_{b}$ by imposing different densities
on two boundaries of the system. From here on, we use square brackets to denote the dependence of the observable on the boundary-condition parameters. Additionally, the notation in Eq.~\eqref{eq:responseJ} implies a tensor product such that $\left(\mathbf{A}\mathbf{B}\right)_{ij}=A_i B_j$.

Eq.~\eqref{eq:Fa-1} can also be seen as the response to a density
perturbation $\delta\rho_{j}\left(\br\right)\equiv\rho_{j}\left(\br\right)-\rho_{b}$; thus
\begin{align}
\mathbf{F}_{12}^{a} & =\mathbf{R}_{2}^{\rho}\,\delta\rho_{1}\!\left(\br_{12}\right)+\mathcal{O}\!\left(r_{12}^{-2}\right),\label{eq:responserho}
\end{align}
where we define the linear response operator
\begin{align} \label{eq:Rjrho}
\mathbf{R}_{j}^{\rho} & \equiv\left.-\partial_{\rho_{b}}\,\mathbf{p}_{j}\!\left[\rho_{b},\bj_{b}\right]\right|_{\bj_{b}=0}=\int d^{2}\br\left.\partial_{\rho_{b}}\,\rho\!\left[\rho_{b},\bj_{b}\right]\right|_{\bj_{b}=0}\bnabla V_{j}.
\end{align}
Note that Eqs.~\eqref{eq:multip},~\eqref{eq:ffrho} and~\eqref{eq:Jff} imply that the forces decay with distance as ${\bf F}_{12}^{a} \sim r_{12}^{-1}$ and ${\bf F}_{12}^{s}\sim r_{12}^{-2}$. Importantly, ${\bf F}_{12}$ can be
predicted, to leading order, solely by measuring the single-body properties
$\bp_{1}$, $\bp_{2}$, ${\bf R}_{2}^{P}$ (or ${\bf R}_{2}^{\rho}$)
and $\mathbb{R}_{2}$. In practice, for bodies with an axis of symmetry, say the $x$ axis, the measurement is reduced even further. By reflection symmetry, the dipole moment satisfies $\bp_2=p_2 \mathbf{e}_x$ (with $p_2$ not necessarily positive). Hence, $\mathbf{R}_2^P=R_2^P \mathbf{e}_x$, meaning that one has to measure only the $x$ component. In a similar manner, one of the principal axes of $\mathbb{R}_2^P$ coincides with the $x$ axis, and therefore the other is the $y$ axis. This allows one to measure only two components of this tensor, instead of four. Note that ${\bf F}_{12}$ is not necessarily
symmetric under the exchange of indices $1\leftrightarrow2$, indicating
that an action-reaction principle for passive bodies interactions which are mediated by active particles
does not hold. This property is expected, because the forces ${\bf F}_{12}$
and ${\bf F}_{21}$ are mediated by dry active particles, which do
not conserve momentum.

The physics of Eqs.~\eqref{eq:Fa-1} and~\eqref{eq:Fs-1} can be described
as follows. To leading order, the influence of body $1$ can be attributed to a local shift of the pressure field $\Delta P_{1}\!\left(\br_{12}\right)$
or to a local shift of the particle density $\delta\rho_{1}\!\left(\br_{12}\right)$.
Since this shift is a scalar quantity, it contributes only to ${\bf F}_{12}^{a}$
and thus can only modulate the force on an asymmetric body $2$. At the next order,
body $1$ also generates a constant current of density $\bj_{1}\!\left(\br_{12}\right)$.
This current applies a force on body $2$ even if it is fully symmetric.
Thus, it provides the leading-order contribution to ${\bf F}_{12}^{s}$,
with $\mathbb{R}_{2}$ being the response operator. Hence, an asymmetric
body $1$ can propel a fully symmetric body $2$ in the direction
of $\bj_{1}$. 

The additional torque exerted on body $2$ due to body $1$, $\bm{\tau}_{12}$,
can be expressed in a similar manner (see Fig.~\ref{fig:interaction}). We denote the self-torque of the
isolated bodies by
\begin{align}
\bm{\tau}_{j} & =\int d^{2}\mathbf{r}\,\rho_{j}\!\left(\mathbf{r}\right)\left(\mathbf{r}-{\bf X}_{j}\right)\times\bm{\nabla}V_{j},\label{eq:st}
\end{align}
where ${\bf X}_{j}$ is the position of body $j$, satisfying ${\bf X}_{1}-{\bf X}_{2}=-\br_{12}$.
Clearly, depending on the shape of the body in question, it may or may not
experience a self-torque. For example, a spherically symmetric body, for which
$V_{j}=V_{j}\left(\abs{\br-{\bf X}_{j}}\right)$, experiences no self-torque.
It has been already demonstrated, both numerically\cite{NikolaPRL2016}
and experimentally\cite{DiLeonardoPNAS2010,SokolovPNAS2010}, that
asymmetric bodies generate ratchet currents that induce
a self-torque.

We identify $\bm{\tau}_{12}$ as the change in the self-torque
of body $2$ due to the introduction of body $1$ into the fluid.
As done for the interaction force, we decompose $\bm{\tau}_{12}$
as
\begin{align}
\bm{\tau}_{12} & =\bm{\tau}_{12}^{a}+\bm{\tau}_{12}^{s},\label{eq:tdec}
\end{align}
where $\bm{\tau}_{12}^{a}$ acts only on bodies with non-zero self-torque
($\bm{\tau}_{2}\neq0$), while $\bm{\tau}_{12}^{s}$ is present even
for bodies with zero self-torque ($\bm{\tau}_{2}=0$). At leading
order in $r_{12}$, 
\begin{align}
\bm{\tau}_{12}^{a} & =\mathbf{T}_{2}\,\Delta P_{1}\!\left(\br_{12}\right)+\mathcal{O}\!\left(r_{12}^{-2}\right),\label{eq:ta}\\
\bm{\tau}_{12}^{s} & =\bm{\gamma}_{2}\times\bj_{1}\!\left(\br_{12}\right)+\mathcal{O}\!\left(r_{12}^{-3}\right),\label{eq:ts}
\end{align}
where we define the linear response operators 
\begin{align}
\mathbf{T}_{j}^{P} & \equiv\left.\partial_{P\!\left(\rho_{b}\right)}\,\bm{\tau}_{j}\!\left[P\!\left(\rho_{b}\right),\bj_{b}\right]\right|_{\bj_{b}=0}=\int d^{2}\br\left.\partial_{P\!\left(\rho_{b}\right)}\,\rho\!\left[P\!\left(\rho_{b}\right),\bj_{b}\right]\right|_{\bj_{b}=0}\br\times\bnabla V_{j},\label{eq:response-2}\\
\bm{\gamma}_{j} & \equiv\left.\bnabla_{\bj_{b}}\times\bm{\tau}_{j}\!\left[P\!\left(\rho_{b}\right),\bj_{b}\right]\right|_{\bj_{b}=0}=-\int d^{2}\br\left.\bnabla_{\bj_{b}}\,\rho\!\left[P\!\left(\rho_{b}\right),\bj_{b}\right]\right|_{\bj_{b}=0}\cdot\br\bnabla V_{j}\;. \label{eq:gamma}
\end{align}
In the last equality we have used the triple product vector identity. In the spirit of Eq.~\eqref{eq:responserho}, Eq.~\eqref{eq:ta} can also be seen as the response to the density perturbation
$\delta\rho_{1}$; thus
\begin{align}
\bm{\tau}_{12}^{a} & =\mathbf{T}_{2}^{\rho}\,\delta\rho_{1}\!\left(\br_{12}\right)+\mathcal{O}\!\left(r_{12}^{-2}\right),\label{eq:Fa-1-1-1}
\end{align}
where we define the linear response operator
\begin{align}
\mathbf{T}_{j}^{\rho} & \equiv\left.-\partial_{\rho_{b}}\,\bm{\tau}_{j}\!\left[\rho_{b},\bj_{b}\right]\right|_{\bj_{b}=0}=\int d^{2}\br\left.\partial_{\rho_{b}}\rho\left[\rho_{b},\bj_{b}\right]\right|_{\bj_{b}=0}\br\times\bnabla V_{j}.\label{eq:trho}
\end{align}

Once more, we find that the interaction torque can be expressed, to
leading order, using measurable single-body properties. For a body with the $x$ axis as an axis of symmetry, we have by reflection symmetry $\mathbf{T}_{j}^{P}=T_j^P \mathbf{e}_x$, $\mathbf{T}_{j}^{P}=T_j^\rho \mathbf{e}_x$ and $\bm{\gamma}_{j}=\gamma_j \mathbf{e}_x$, which reduces the number of components required for measurement to two, instead of four. As was the case for the interaction force, because the local shift of the pressure
field $\Delta P_{1}\!\left(\br_{12}\right)$ (or equivalently, the local
shift of the particle density $\delta\rho_{1}\!\left(\br_{12}\right)$)
is a scalar quantity, it contributes only to $\bm{\tau}_{12}^{a}$. Thus
it can modify only an already-existing self-torque on body $2$ about its axis, but not
generate a torque by itself. In contrast, due to the non-uniform flow in the
vicinity of the body, the local current density $\bj_{1}\!\left(\br_{12}\right)$
can exert a torque on body $1$ even if it has zero self-torque. Hence,
an asymmetric body $1$ can cause body $2$ to rotate,
even if it has no self-torque. As seen in Eq.~\eqref{eq:ts}, this rotation
tends to align $\bm{\gamma}_{2}$, a body dependent quantity, with ${\bf J}_{1}\!\left(\br_{12}\right)$.

Finally, we note that all the results summarized in this section reproduce the non-interacting case derived in Ref. \cite{BaekPRL2018}; this can be easily checked using the equation of state for the "classical active gas" $P\!\left(\rho_{b}\right)=T_\text{eff}\,\rho_{b}$, for which ${\bf R}_{j}^{P}=\bp_{j}/(T_\text{eff}\,\rho_{b})$
and ${\bf R}_{j}^{\rho}=\bp_{j}/\rho_{b}$. We next proceed to detailed derivations of the above results.

\section{Steady-state equations}
\label{sec:hydrodynamics}

In order to obtain the above results, we use the steady-state equations for empirical distributions averaged over histories. These can be derived directly from the particle dynamics described in Sec.~\ref{sec:model} using standard methods \cite{DeanJPA1996}. This was carried out in Refs. \cite{NikolaPRL2016,SolonNJP2018} for the case of pairwise interacting ABPs ($\alpha=0$) and in Ref. \cite{BaekPRL2018} for the case of non-interacting particles ($U=0$). Here we use these references and present the equations in their general form. We also outline the explicit derivation for the case $\alpha=0$ in Appendix \ref{sec:wie}.

We are interested in the marginal empirical distributions
\begin{align}
\hat{\mathbf{m}}^{\left(n\right)}\!\left(\mathbf{r}\right)\equiv\sum_{i}\delta\!\left(\mathbf{r}-\mathbf{r}_{i}\right)\mathbf{e}_{n\theta_{i}} \label{eq:marg}
\end{align}
for integers $n \geq 0$, where $\mathbf{e}_{n \theta}=\left(\cos\left(n \theta\right),\sin\left(n \theta\right)\right)^{T}$ is the $n$th harmonic unit vector. In particular, we have the identities $\hat{\mathbf{m}}^{\left(0\right)}= \left(\hat{\rho},\,0\right)^{T}$
and $\hat{\mathbf{m}}^{\left(1\right)}=\hat{\mathbf{m}}$. Taking an average over histories, $\mb^{(n)}\!\left(\mathbf{r},\theta\right) \equiv \braket{\hat{\mb}^{(n)}\!\left(\mathbf{r},\theta\right)}$, and considering the steady state, where $\partial_{t} \mb^{(n)}=0$, one obtains for the special case $n = 0$ a zero-flux condition,
\begin{align}
\bm{\nabla}\cdot\bj=0,\label{eq:cont}
\end{align}
with the current density given by~\cite{SolonNJP2018}
\begin{align}
\mathbf{J} & =-\mu\rho\bnabla V+\mu l_\text{r}\bnabla\cdot\left[\left(\bm{\nabla}V\right)\mathbf{m}\right]+\mu\bnabla\cdot\bm{\sigma}.\label{eq:cdyn-1}
\end{align}
Here $\bm{\sigma}$ is the stress tensor given by
\begin{align}
\bm{\sigma} & =-T_\text{eff}\,\rho\,\id+\bm{\sigma}^\text{IK}+\bm{\sigma}^\text{P}.\label{eq:sigma}
\end{align}
In this decomposition, the ideal gas component is $-T_\text{eff}\,\rho\,\id$,
and the polarization component $\bm{\sigma}^\text{P}$ is given by 
\begin{align}
\bm{\sigma}^\text{P}\!\left(\br\right) & =l_\text{r}\int d^{2}\mathbf{r}'\left[\bnabla U\!\left(|\mathbf{r}-\mathbf{r}'|\right)\right]\braket{\hat{\mathbf{m}}\!\left(\br\right)\hat{\rho}\!\left(\mathbf{r}'\right)}+Tl_\text{r}\bm{\nabla}\mathbf{m}\!\left(\br\right)-2\left(T_\text{eff}-T\right)\mathbb{Q}\!\left(\br\right),\label{eq:sigmap}
\end{align}
where $T \equiv D_\text{t}/\mu$ denotes the temperature of the ambient thermal bath and \begin{align}
\hat{\mathbb{Q}}\!\left(\br\right)\equiv\sum_{i}\delta\!\left(\mathbf{r}-\mathbf{r}_{i}\right)\left(\mathbf{e}_{\theta_{i}}\mathbf{e}_{\theta_{i}}-\frac{1}{2}\id\right) = \frac{1}{2}\begin{pmatrix}\hat{m}_x^{(2)} & \hat{m}_y^{(2)}\\
\hat{m}_y^{(2)} & -\hat{m}_x^{(2)}\end{pmatrix},\label{eq:nem}
\end{align}
with $\id$ denoting the identity tensor, is the nematic order tensor.
The interaction component $\bm{\sigma}^\text{IK}$ satisfies 
\begin{align}
\bnabla\cdot\bm{\sigma}^\text{IK}\!\left(\br\right) & =-\int d^{2}\mathbf{r}'\braket{\hat{\rho}\!\left(\mathbf{r}\right)\hat{\rho}\!\left(\mathbf{r}'\right)}\bnabla U\!\left(|\mathbf{r}-\mathbf{r}'|\right),\label{eq:int}
\end{align}
and is given by the standard Irving-Kirkwood formula~\cite{IrvingJChem1950,KrugerJChem2018}
\begin{align}
\bm{\sigma}^\text{IK}\!\left(\br\right) & =\frac{1}{2}\int d^{2}\br'\frac{\br'\br'}{r'}U'\!\left(r'\right)\int_{0}^{1}d\lambda\braket{\hat{\rho}\!\left(\br+\left(1-\lambda\right)\br'\right)\hat{\rho}\!\left(\br-\lambda\br'\right)}.\label{eq:IK}
\end{align}
We note that Eqs.~\eqref{eq:pd} and \eqref{eq:pi} are obtained from Eqs.~\eqref{eq:sigmap} and \eqref{eq:IK} by $P_\text{D} = -\tr\bm{\sigma}^\text{IK}/2$ and $P_\text{I}=-\tr\bm{\sigma}^\text{P}/2$.

In addition, for $n \ge 1$, one obtains~\cite{NikolaPRL2016,BaekPRL2018}
\begin{align}
\mb^{\left(n\right)}\!\left(\br\right) & =-\frac{1}{\alpha+n^{2}D_\text{r}}\bm{\nabla}\cdot\left\{-\mu\left[\bm{\nabla}V\!\left(\br\right)\right]\mb^{\left(n\right)}\!\left(\br\right)-\mu\int d^{2}\mathbf{r}'\left[\bnabla U\!\left(|\mathbf{r}-\mathbf{r}'|\right)\right]\braket{\hat{\mathbf{m}}^{\left(n\right)}\!\left(\br\right)\hat{\rho}\!\left(\mathbf{r}'\right)}-D_\text{t}\bm{\nabla}\mb^{\left(n\right)}\!\left(\br\right)\right\}\nonumber \\
 & -\frac{v}{2\left(\alpha+n^{2}D_\text{r}\right)}\left[\mathbb{D}\mb^{\left(n-1\right)}\!\left(\br\right)-\mathbb{D}^{\dagger}\mb^{\left(n+1\right)}\!\left(\br\right)\right],\label{eq:deanmn-2-1}
\end{align}
where $\mathbb{D}$ and $\mathbb{D}^\dagger$ are the antisymmetric roots of $-\nabla^{2}$, defined as
\begin{align}
\mathbb{D}\equiv\begin{pmatrix}\partial_{x} & -\partial_{y}\\
\partial_{y} & \partial_{x}
\end{pmatrix}, & \quad\mathbb{D}^{\dagger}\equiv\begin{pmatrix}-\partial_{x} & -\partial_{y}\\
\partial_{y} & -\partial_{x}
\end{pmatrix}.
\end{align}
For $n=1$, Eq.~\eqref{eq:deanmn-2-1} implies~\cite{SolonNJP2018}
\begin{align}
v\mb\!\left(\br\right) & =-l_\text{r}\bm{\nabla}\cdot\left\{-\mu\left[\bm{\nabla}V\!\left(\br\right)\right]\mathbf{m}\!\left(\br\right)-\mu\int d^{2}\mathbf{r}'\left[\bnabla U\!\left(|\mathbf{r}-\mathbf{r}'|\right)\right]\braket{\hat{\mathbf{m}}\!\left(\br\right)\hat{\rho}\!\left(\mathbf{r}'\right)}-D_\text{t}\bm{\nabla}\mathbf{m}\!\left(\br\right)+v\mathbb{Q}\!\left(\br\right)\right\} -\frac{vl_\text{r}}{2}\bnabla\rho\!\left(\br\right),\label{eq:vm}
\end{align}
with $\mathbb{Q}$ satisfying $\bnabla\cdot\hat{\mathbb{Q}}=-\mathbb{D}^{\dagger}\hat{\mb}^{\left(2\right)}/2$.

\section{Far-field effects of a single body}
\label{sec:far-field}

We next consider a single passive body immersed in a homogeneous
active fluid of density $\rho_{b}$. The body is described by
the potential $V$, which is zero beyond a finite distance $\sim d$.
The diameter $d$ and the run-length $l_\text{r}$ define two microscopic length scales.
In the following, we derive the pressure, density, and current perturbation fields in the far-field limit $r\gg \max\left(l_\text{r},d\right)$.
We later build on these to derive the interactions between two bodies mediated by the active fluid.

\subsection{Pressure field}

We first derive Eq.~\eqref{eq:multip}, which describes the far-field behavior of the pressure field. Toward this goal, we examine the standard deviatoric decomposition
\begin{align}
\bm{\sigma} & =-P\id+\mathbb{S},\label{eq:dev}
\end{align}
where $P\equiv-\tr\bm{\sigma}/2$ is the pressure field, and $\mathbb{S}\equiv\bm{\sigma}-\id\tr\bm{\sigma}/2$
is the traceless deviatoric stress tensor. We can also represent $\bnabla\cdot\bm{\sigma}$, which is a vector, using the Helmholtz decomposition
\begin{align}
\bnabla\cdot\bm{\sigma} & =-\bnabla\Phi+\bnabla\times\bm{\Psi},\label{eq:helm}
\end{align}
where $\Phi$ and $\bm{\Psi}$ are scalar and vector potentials, respectively. Similarly, we can write the Helmholtz decomposition
of $\bnabla\cdot\mathbb{S}$,
\begin{align}
\bnabla\cdot\mathbb{S} & =-\bnabla\Phi_{S}+\bnabla\times\bm{\Psi},\label{eq:helm-1}
\end{align}
where $\Phi_{S}$ is the corresponding scalar potential. It is clear that the same vector potential $\bm{\Psi}$ can be used in both decompositions because $\bm\sigma$ and $\mathbb{S}$ differ only by a scalar multiple of $\id$. Indeed, one can easily check that Eqs.~\eqref{eq:dev}, \eqref{eq:helm} and \eqref{eq:helm-1} are mutually consistent if the scalar potentials are related by
\begin{align}
\Phi & =P+\Phi_{S}.\label{eq:scalars}
\end{align}
We can interpret the above relations as follows: (1) the shear stress $\mathbb{S}$ contributes to the scalar potential $\Phi_{S}$ and the vector potential $\Psi$; (2) the shear stress $\mathbb{S}$ also contributes to the scalar potential $\Phi$ via $\Phi_{S}$. With this structure in mind, we proceed by first calculating the far-field behavior of $\Phi$ and then showing that the shear-stress component $\Phi_{S}$ is negligible as it contributes only to higher-order corrections.

Taking the divergence of Eq.~\eqref{eq:cdyn-1} and using
the steady-state condition $\bnabla\cdot\bj=0$, one gets
\begin{align}
\partial_{\alpha}\partial_{\beta}\sigma_{\alpha\beta} & =\bnabla\cdot\left(\rho\bnabla V\right)-l_\text{r}\partial_{\alpha}\partial_{\beta}\left(m_{\alpha}\partial_{\beta}V\right).\label{eq:divsig}
\end{align}
On the other hand, taking the divergence of Eq.~\eqref{eq:helm} gives
\begin{align}
\lap\Phi & =-\partial_{\alpha}\partial_{\beta}\sigma_{\alpha\beta}.
\label{eq:lapphi}
\end{align}
Combining these two equations, we obtain the Poisson equation
\begin{align}
\lap\Phi & =-\bnabla\cdot\left(\rho\bnabla V\right)+l_\text{r}\partial_{\alpha}\partial_{\beta}\left(m_{\alpha}\partial_{\beta}V\right).\label{eq:pois}
\end{align}
To solve this equation by the method of Green's functions, one should clarify the boundary conditions at infinity. These are fixed by assuming that the active fluid is homogeneous and disordered at $r\rightarrow\infty$. Since there is no preferred direction, $\bm{\sigma}\!\left(\rho_{b}\right)=\lim_{r\to\infty}\bm{\sigma}\!\left(\br\right)$ is isotropic -- there is no spontaneous shear at $r\rightarrow\infty$. Then, by the deviatoric decomposition~\eqref{eq:dev}, $\sigma = -P \id$ and $\mathbb{S} = 0$ for $r\rightarrow\infty$, which in turn implies $\Phi_{S} = 0$ in this limit. Thus the boundary condition at infinity is obtained as $\lim_{r\to\infty}\Phi(\mathbf{r}) = P(\rho_b)$. It should be noted that this result relies on the spherical symmetry of the interaction potential $U$; without this symmetry, $\bm{\sigma}\!\left(\rho_{b}\right)$ generally depends on the correlations among $\hat{\mathbf{m}}^{(n)}$ with $n \geq 1$.

Based on this boundary condition, Eq.~\eqref{eq:pois} is solved by
\begin{align}
\Phi\!\left(\br\right) & = P\!\left(\rho_{b}\right)-\frac{1}{2\pi}\int d^{2}\mathbf{r}'\ln\left|\mathbf{r}-\mathbf{r}'\right|\left\{ \bnabla'\cdot\left[\rho\!\left(\br'\right)\bnabla'V\!\left(\br'\right)\right]-l_\text{r}\partial_{\alpha}'\partial_{\beta}'\left[m_{\alpha}\left(\br'\right)\partial_{\beta}'V\!\left(\br'\right)\right]\right\}\;, \label{eq:green}
\end{align}
where $\bnabla' = \bnabla_{\bf r'}$. Taking a multipole expansion, we
obtain
\begin{align}
\Phi\!\left(\br\right) & = P\!\left(\rho_{b}\right)+\frac{1}{2\pi}\frac{\mathbf{r}\cdot\mathbf{p}}{r^{2}}+\mo{r^{-2}},\label{eq:multip-2}
\end{align}
where the dipole moment $\mathbf{p}$ is as defined in Eq.~\eqref{eq:dipole-1-1}. We stress that the above formula relies on the assumption of a homogeneous and disordered fluid with a symmetric pairwise potential.

To obtain the far-field behavior of the pressure field from Eq.~\eqref{eq:multip-2}, we need information about the far-field behavior of $\Phi_{S}$. In general, from Eqs.~\eqref{eq:sigma}, \eqref{eq:sigmap}, and \eqref{eq:IK}, $\bm{\sigma}$ can be expressed as a local function of $\rho$, $\mb^{\left(n\right)}$, $\braket{\hat{\rho}^2}$, and $\braket{\hat{\mb}^{\left(n\right)}\,\hat{\rho}}$ with $n \ge 1$, and their spatial derivatives. However, we expect that the far-field behavior of $\bm{\sigma}$ would be dominated by the local contribution from $\rho$, so that one can write
\begin{align}
\bm{\sigma}(\mathbf{r}) & =\bm{\sigma}\!\left(\rho(\mathbf{r})\right)+\mo{\partial\rho} \;,\label{eq:sigasum}
\end{align}
where $\partial$ stands for a spatial derivative. This can be justified mathematically based on two assumptions: (1) $U$ is short-ranged; (2) $U$ is weak or pair correlations of the empirical densities satisfy mean-field properties---see Appendix~\ref{sec:stress} for the detailed derivation. More importantly, as we show below, the results derived from Eq.~\eqref{eq:sigasum} are consistent with our numerical simulation.

With Eq.~\eqref{eq:sigasum}, we proceed by taking a Taylor expansion
\begin{align}
\bm\sigma\!\left(\rho\right) & =-P\!\left(\rho_{b}\right)\id-\left(\rho-\rho_{b}\right)P'\!\left(\rho_{b}\right)\id+\mathcal{O}\left[\left(\rho-\rho_{b}\right)^{2}\right],\label{eq:sigexp0}
\end{align}
which shows that the components of the deviatoric decomposition~\eqref{eq:dev} satisfy
\begin{align}
P &= P\!\left(\rho_{b}\right)+\left(\rho-\rho_{b}\right)P'\!\left(\rho_{b}\right)+\mathcal{O}\!\left[\left(\rho-\rho_{b}\right)^{2},\partial\rho\right],\label{eq:sigexp1}\\
\mathbb{S} &= \mathcal{O}\!\left[\left(\rho-\rho_{b}\right)^{2},\partial\rho\right],\label{eq:sigexp2}
\end{align}
in the far field. From Eq.~\eqref{eq:sigexp1}, one observes that the far-field pressure satisfies $P-P\!\left(\rho_{b}\right)\sim\rho-\rho_{b}$
and $\partial P\sim\partial\rho$. Thus Eq.~\eqref{eq:sigexp2}
can be rewritten as $\mathbb{S}=\mo{ \left[P-P\!\left(\rho_{b}\right)\right]^2 ,\partial P}$, implying
\begin{align}
\bnabla\cdot\mathbb{S} = \mathcal{O}\!\left( \left[P-P\!\left(\rho_{b}\right)\right]\partial P,\,\partial^{2}P\right).\label{eq:Sdeco}
\end{align}
Note that a posteriori we expect $ P-P\!\left(\rho_{b}\right) \sim r^{-1}$, meaning that $\mathbb{S}= \mathcal{O}\!\left( \partial P \right)=\mathcal{O}\!\left(r^{-2} \right)$ and $\bnabla\cdot\mathbb{S}= \mathcal{O}\!\left( \partial^{2}P \right)=\mathcal{O}\!\left(r^{-3} \right)$.
In general, $\Phi_{S}$ and $\bm{\Psi}$ satisfying the Helmholtz decomposition~\eqref{eq:helm-1} are not local functions of $\bnabla\cdot\mathbb{S}$, so the relation between the far-field behaviors of $\Phi_{S}$ and $\mathbb{S}$ is not immediately obvious. However, as discussed in Appendix \ref{sec:Scalar-and-vector}, we can show that the far-field behaviors of both $\Phi_S$ and $\bm\Psi$ are of order $\mathcal{O}\!\left(\mathbb{S},r^{-2}\right)$. In other words, $\Phi_{S} = \mathcal{O}\!\left(\left[P-P\!\left(\rho_{b}\right)\right]^{2},\partial P,r^{-2}\right)$, so Eq.~\eqref{eq:scalars} yields a far-field approximation
\begin{align}
\Phi = P+\mathcal{O}\!\left(\left[P-P\!\left(\rho_{b}\right)\right]^{2},\partial P,r^{-2}\right),\label{eq:phiP}
\end{align}
which justifies writing $\Phi-P\!\left(\rho_{b}\right)\sim P-P\!\left(\rho_{b}\right)$ and $\partial\Phi\sim\partial P$.
Then we can invert Eq.~\eqref{eq:phiP} to obtain
\begin{align}
P & =\Phi+\mathcal{O}\!\left(\left[\Phi-P\!\left(\rho_{b}\right)\right]^{2},\partial\Phi,r^{-2}\right).\label{eq:Pphi}
\end{align}
Using Eq.~\eqref{eq:multip-2} in the above relation, we finally obtain the far-field expression for the pressure field shown in Eq.~\eqref{eq:multip}.

\subsection{Density and current fields}

To obtain the far-field expressions for $\rho$ and $\bj$, we first
note that the particle density and the pressure field are related in the far-field by $P-P\!\left(\rho_{b}\right)\sim\rho-\rho_{b}$
and $\partial P\sim\partial\rho$. Using these, Eq.~\eqref{eq:sigexp1} can be inverted as
\begin{align}
\rho & =\rho_{b}+\frac{P-P\!\left(\rho_{b}\right)}{P
'\!\left(\rho_{b}\right)}+\mathcal{O}\left\{ \left[P-P\!\left(\rho_{b}\right)\right]^{2},\partial P\right\} .\label{eq:rhoP}
\end{align}
Substituting Eq.~\eqref{eq:multip} into Eq.~\eqref{eq:rhoP}, we
obtain the multipole expansion for $\rho$ shown in Eq.~\eqref{eq:ffrho}.

We now turn to the far-field current density. Substituting Eq.~\eqref{eq:multip} into Eqs.~\eqref{eq:sigasum} and \eqref{eq:sigexp0} and noting that $\left(\rho-\rho_{b}\right)^{2}=\mo{r^{-2}}$, we find 
\begin{align}
\bm{\sigma}(\mathbf{r}) =\left[P\!\left(\rho_{b}\right)-\frac{1}{2\pi}\frac{\mathbf{r}\cdot\mathbf{p}}{r^{2}}\right]\id+\mo{r^{-2}}.\label{eq:sigexp-1}
\end{align}
Substituting Eq.~\eqref{eq:sigexp-1} into Eq.~\eqref{eq:cdyn-1},
and using the fact that outside the body Eq.~\eqref{eq:cdyn-1}
becomes $\bj=\mu\bnabla\cdot\bm{\sigma}$, we obtain Eq.~\eqref{eq:Jff}.
This means that, up to $\mo{r^{-3}}$, $\bj$ is curl-free and behaves like the gradient of a scalar potential $\mu P$, with $\rot\bj=\mo{r^{-4}}$.

\begin{figure}
  \centering
  \includegraphics[width=0.53\textwidth]{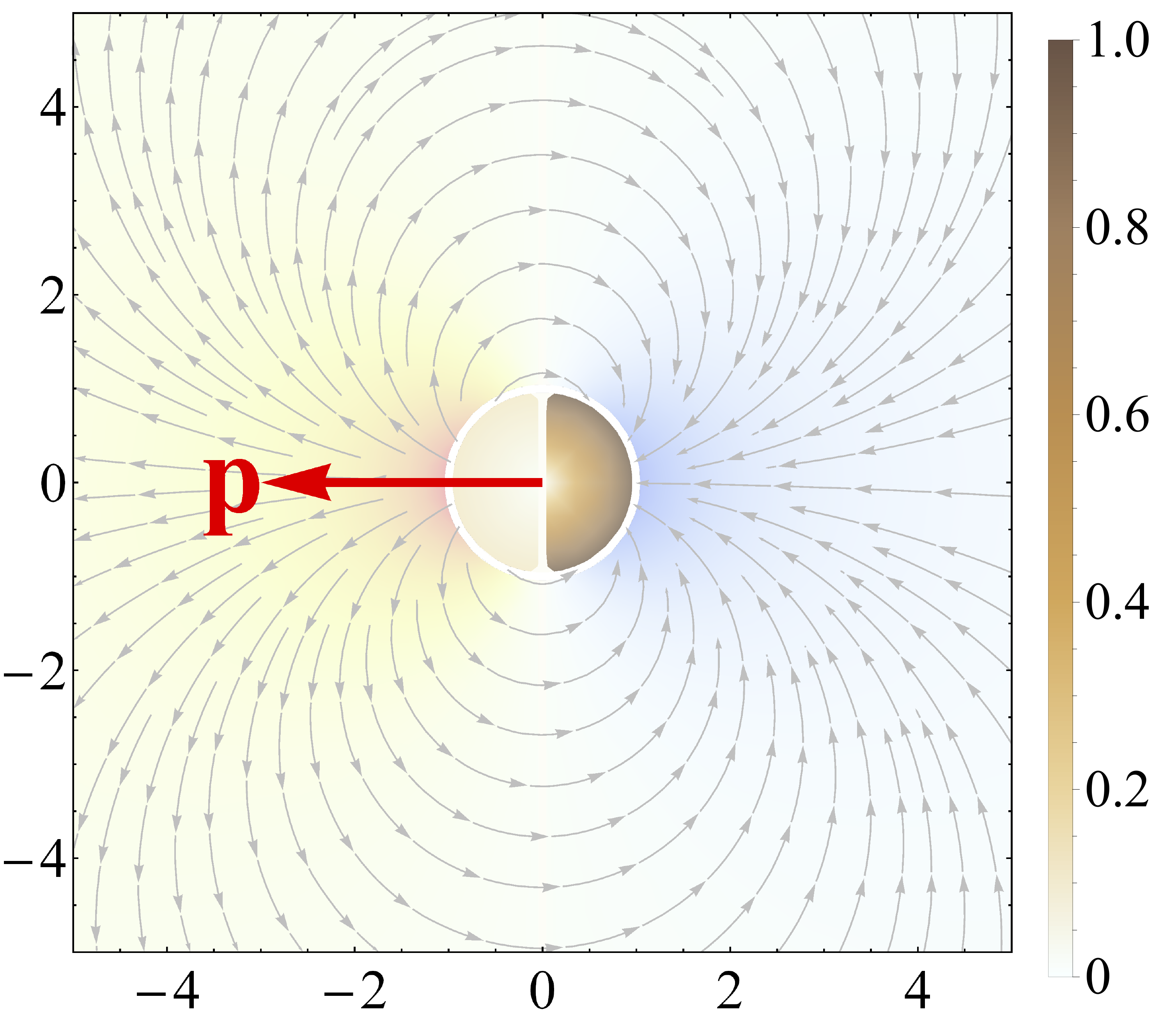}
  \caption{The simulated potential (brown colors), representing an asymmetric passive body. Theoretical prediction of a dipolar current density $\bj\left(\br\right)$ according to Eq.~\eqref{eq:Jff} is shown in gray streamlines. Prediction for dipolar density and pressure perturbations according to Eq.~\eqref{eq:multip} and~\eqref{eq:ffrho} is shown in red and blue map. The dipole moment $\bp$ is drawn schematically in red in the negative $x$-direction.}
  \label{fig:diagram}
\end{figure}

\begin{figure}[h]
  \centering
  \includegraphics[width=0.7\textwidth]{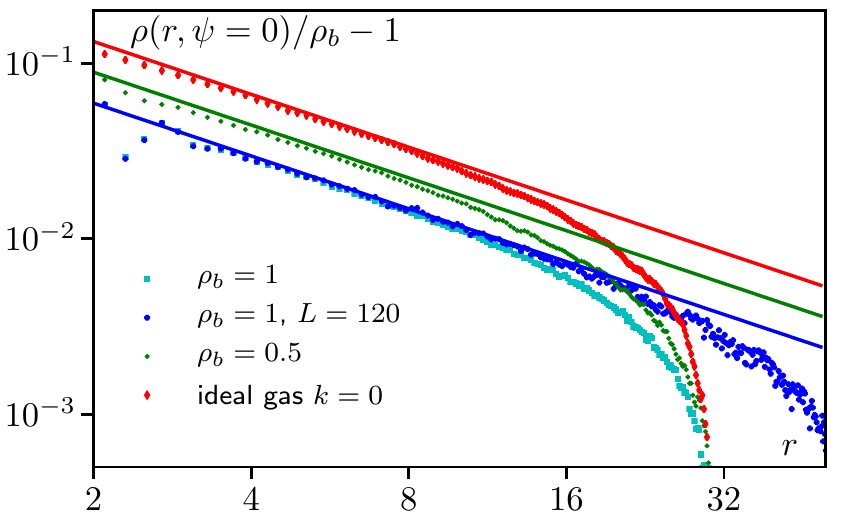}
  \includegraphics[width=0.7\textwidth]{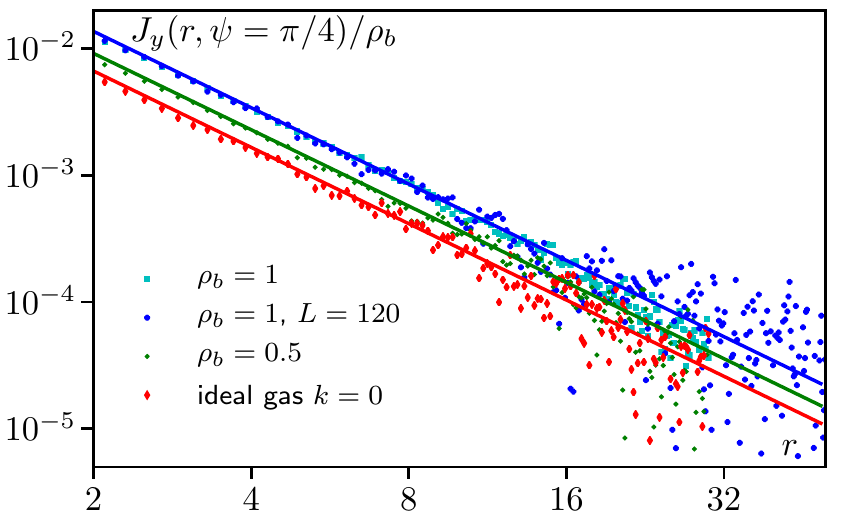}
  \caption{Comparison between the density (top) and current (bottom)
    measured in numerical simulations (symbols) and the predictions
    from Eqs.~\eqref{eq:ffrho} and \eqref{eq:Jff} (solid
    lines). Parameters: $v=1$, $\alpha=5$, $D_\mathrm{r}=0$, $k=2$, system size
    $L=60$, $dt=0.01$ unless otherwise noted.}
  \label{fig:numerics-dens-current}
\end{figure}

The above result relies on the assumption made in Eq.~\eqref{eq:sigasum} that the stress tensor at the leading order can be expressed as a function of the local density. To verify this, we numerically check the density and current fields predicted by Eqs.~\eqref{eq:ffrho} and \eqref{eq:Jff} using a molecular dynamics simulation. For the simulation, we consider particles interacting
through a short-ranged harmonic repulsion, taking
$U(r)=\frac{k}{2}(1-r)^2$ if $r<1$ and $U(r)=0$ otherwise as the interaction potential in
Eq.~\eqref{eq:langevinr}. For the external potential describing the body--particle interaction, we choose an asymmetric
repulsive potential, taking $V({\bf r})=a(x) r$ if $r<1$ and $V=0$ otherwise. The
coefficient $a(x)$ controls the asymmetry of the object (see Fig.~\ref{fig:diagram}). We take
$a(x)=0.9$ if $x>0$ and $a(x)=0.1$ if $x<0$, with the other
parameters set to be $v=1$, $k=2$, $\alpha=5$, and
$D_\text{r}=0$. Numerical integrations of Eqs.~\eqref{eq:langevinr} and \eqref{eq:langevint} are carried out
using Euler's method with a discrete time step $dt=0.01$. To compute the compressibility
$c(\rho_b)$ appearing in Eq.~(\ref{eq:ffrho}), we first independently measure
the pressure as a function of density in the absence of the body. Then, after adding the
body, the dipole moment ${\bf p}$ is measured from
Eq.~\eqref{eq:dipole-1-1} for different values of the density $\rho_b$ of active
particles. Given the symmetry of the problem, the dipole moment should be parallel to the
$x$-axis. Finally, we compare the measured density and current fields
to the theoretical prediction of Eqs.~\eqref{eq:ffrho} and \eqref{eq:Jff}. The two show an excellent agreement {\it without any fitting parameters}. Two
examples are shown in Fig.~\ref{fig:numerics-dens-current}: the
density field along the $x$-axis (at $\psi\equiv \arg {\bf r}=0$) and the
$y$-component of the current at $\psi=\pi/4$. For $\rho_b=1$, we
display data for a larger system with $L=120$ to show that the discrepancy
at large $r$ for the density field is a finite-size effect.

It is interesting to note that we see two opposing trends in our numerical example. The dipole moment
increases superlinearly with the density of active particles,
so that the normalized current $|{\bf J}/\rho_b|$ increases with
density $\rho_b$ in Fig.~\ref{fig:numerics-dens-current}. On the contrary, the normalized
disturbance in the density field $\left(\rho-\rho_b\right)/\rho_b$ decreases with $\rho_b$ because of the
decrease of the compressibility.

\newpage
\newpage

Finally, in Appendix \ref{sec:Finite-Size-Effects}, we show that by carefully taking the infinite system limit, one recovers the previously derived
current-force relation\cite{NikolaPRL2016,BaekPRL2018}
\begin{align}
\int d^{2}r\,\mathbf{J}\left(\br\right) & =\mu\bp.\label{eq:cfr}
\end{align}
Moreover, we show that for periodic systems of size $\sim L$, the
correction to the particle density decays with $L$ as $\mathcal{O}\left(L^{-2}\right)$.

\section{Long-range interactions between bodies}
\label{sec:interactions}

We now consider a pair of static bodies fixed in a fluid of density
$\rho_{b}$ and infinite volume. Interactions of the bodies with active
particles are described by two potentials $V_{1}$ and $V_{2}$, each localized in space. Without
loss of generality, body $2$ is positioned at the origin, and body
$1$ is located at $-\mathbf{r}_{12}$, so that each experiences the far-field effects of the other.

\subsection{Force}

We are interested in the force $\mathbf{F}_{12}$ applied by body
$1$, via the fluid, on body $2$.
More precisely, ${\bf F}_{12}$ is the additional force exerted on body
$2$ by the active particles due to the introduction of body $1$
into the fluid.
To simplify notation, we use a tilde above a quantity to indicate that the value of the quantity
has been modified by the presence of multiple bodies.
For example, the modified force applied on body $j$ can be denoted as $-\tilde{\bp}_{j}$,
where $\tilde{\bp}_{j}\equiv-\int d^{2}\br\,\tilde{\rho}\!\left(\br\right)\bnabla V_{j}\!\left(\br\right)$.
With these notations, the meaning of the interaction force can be expressed concisely by ${\bf F}_{12}\equiv\bp_{2}-\tilde{\bp}_{2}$.

We obtain ${\bf F}_{12}$ by taking a far-field expansion of the contribution of body $1$ to the pressure
field. Since the steady-state conditions derived in Sec.~\ref{sec:hydrodynamics} are valid for arbitrary $V$, we can use the two-body potential $V=V_{1}+V_{2}$ in Eq.~\eqref{eq:green}, which is a direct consequence of the steady-state condition $\bm\nabla\cdot \bj = 0$. Thus Eq.~\eqref{eq:green} can be rewritten in the form of a decomposition
\begin{align}
\Phi & =P\!\left(\rho_{b}\right)+\Delta\tilde{\Phi}_{1}+\Delta\tilde{\Phi}_{2},\label{eq:Pdect}
\end{align}
where 
\begin{align}
\Delta\tilde{\Phi}_{j}\!\left(\br\right) & \equiv-\frac{1}{2\pi}\int d^{2}\mathbf{r}'\ln\left|\mathbf{r}-\mathbf{r}'\right|\left\{ \bnabla'\cdot\left[\tilde{\rho}\!\left(\br'\right)\bnabla'V_{j}\left(\br'\right)\right]-l_\text{r}\partial_{\alpha}'\partial_{\beta}'\left[\tilde{m}_{\alpha}\!\left(\br'\right)\partial_{\beta}'V_{j}\left(\br'\right)\right]\right\}\label{eq:dPj}
\end{align}
accounts for the contribution from body~$j$.
Regarding body $1$ as a far-field object, $\Delta\tilde{\Phi}_{1}$ can be expanded as
\begin{align}
\Delta\tilde{\Phi}_{1} & =\frac{1}{2\pi}\frac{\left(\mathbf{r}_{12}+\mathbf{r}\right)\cdot\tilde{\mathbf{p}}_{1}}{\left|\mathbf{r}_{12}+\mathbf{r}\right|^{2}}+\mathcal{O}\!\left(r_{12}^{-2}\right)\nonumber \\
 & =\frac{1}{2\pi}\left[\frac{\mathbf{r}_{12}\cdot\tilde{\mathbf{p}}_{1}}{r_{12}^{2}}+\frac{\mathbf{r}\cdot\tilde{\mathbf{p}}_{1}}{r_{12}^{2}}-\frac{2\left(\mathbf{r}\cdot\mathbf{r}_{12}\right)\left(\mathbf{r}_{12}\cdot\tilde{\mathbf{p}}_{1}\right)}{r_{12}^{4}}\right]+\mathcal{O}\!\left(r_{12}^{-2},r_{12}^{-3}r^{2}\right)\nonumber \\
 & =\Delta\tilde{P}_{b}+\frac{\tilde{\bj}_{b}\cdot\br}{\mu}+\mathcal{O}\!\left(r_{12}^{-2},r_{12}^{-3}r^{2}\right),\label{eq:multip-1-1-1}
\end{align}
where 
\begin{align}
\Delta\tilde{P}_{b} & \equiv\frac{1}{2\pi}\frac{\mathbf{r}_{12}\cdot\tilde{\mathbf{p}}_{1}}{r_{12}^{2}},\label{eq:dPb}\\
\tilde{\bj}_{b} & \equiv-\frac{\mu}{2\pi}\left[\frac{\tilde{\mathbf{p}}_{1}}{r_{12}^{2}}-\frac{2\left(\mathbf{r}_{12}\cdot\tilde{\mathbf{p}}_{1}\right)\mathbf{r}_{12}}{r_{12}^{4}}\right]\label{eq:dJb}
\end{align}
are the total pressure shift and the current induced by the presence of two bodies in the fluid. Using Eq.~\eqref{eq:multip-1-1-1} in Eq.~\eqref{eq:Pdect}, we get
\begin{align}
\Phi & =P\!\left(\rho_{b}\right)+\Delta\tilde{P}_{b}+\frac{\tilde{\bj}_{b}\cdot\br}{\mu}+\Delta\tilde{P}_{2}+\mathcal{O}\!\left(r_{12}^{-2},r_{12}^{-3}r^{2}\right).\label{eq:Pdect-1}
\end{align}
$\Delta\tilde{P}_{b}$ can be interpreted as a shift in the
pressure around body $2$, and $\tilde{\bj}_{b}\cdot\br/\mu$
as the pressure gradient across body $2$ consistent with the current $\tilde{\bj}_{b}$. Next, we note that $\Delta\tilde{\Phi}_{1}$
can be expressed in terms of single-body properties by expanding $\Phi\!\left[P\!\left(\rho_{b}\right)+\Delta\tilde{P}_{b},\tilde{\bj}_{b}\right]$
with respect to $\Delta\tilde{P}_{b}$
and $\tilde{\bj}_{b}$. The expansion yields
\begin{align}
\Phi & =\left[1+\Delta\tilde{P}_{b}\,\partial_{P\!\left(\rho_{b}\right)}+\tilde{\bj}_{b}\cdot\bnabla_{\bj_{b}}\right]\left.\Phi\!\left[P\!\left(\rho_{b}\right),\bj_{b}\right]\right|_{\bj_{b}=0}+\mathcal{O}\!\left(\Delta\tilde{P}_{b}^{2},\tilde{\bj}_{b}^{2},\tilde{\bj}_{b}\Delta\tilde{P}_{b},r_{12}^{-2}\right)\nonumber \\
 & =\left[1+\Delta\tilde{P}_{b}\,\partial_{P\!\left(\rho_{b}\right)}+\tilde{\bj}_{b}\cdot\bnabla_{\bj_{b}}\right]\left.\Phi\!\left[P\!\left(\rho_{b}\right),\bj_{b}\right]\right|_{\bj_{b}=0}+\mathcal{O}\!\left(r_{12}^{-2}\right),\label{eq:Pexp}
\end{align}
where we have used $\Delta\tilde{P}_{b}=\mathcal{O}\!\left(r_{12}^{-1}\right)$
and $\tilde{\bj}_{b}=\mathcal{O}\!\left(r_{12}^{-2}\right)$. Using the multipole expansion given by Eq.~\eqref{eq:multip-2} on both sides of the equation, we find
\begin{align}
\tilde{\bp}_{2} & =\bp_{2}-\mathbf{R}_{2}^{P}\Delta\tilde{P}_{b}-\mathbb{R}_{2}\tilde{\bj}_{b}+\mathcal{O}\!\left(r_{12}^{-2}\right).\label{eq:pexp}
\end{align}
This implies $\tilde{\bp}_{2}=\bp_{2}+\mo{r_{12}^{-1}}$, which in turn implies $\tilde{\bp}_{1}=\bp_{1}+\mo{r_{12}^{-1}}$ after exchanging the indices $1\leftrightarrow 2$. Then, substituting this back into Eqs.~\eqref{eq:dPb} and~\eqref{eq:dJb},
we obtain
\begin{align}
\Delta\tilde{P}_{b} & =\frac{1}{2\pi}\frac{\mathbf{r}_{12}\cdot\bp_{1}}{r_{12}^{2}}+\mathcal{O}\!\left(r_{12}^{-2}\right)
 =\Delta P_{1}\!\left(\br_{12}\right)+\mathcal{O}\!\left(r_{12}^{-2}\right),\label{eq:dPb-2}\\
\tilde{\bj}_{b} & =-\frac{\mu}{2\pi}\left[\frac{\bp_{1}}{r_{12}^{2}}-\frac{2\left(\mathbf{r}_{12}\cdot\bp_{1}\right)\mathbf{r}_{12}}{r_{12}^{4}}\right]+\mathcal{O}\!\left(r_{12}^{-3}\right).\label{eq:dJb2}
\end{align}
Here $\Delta P_1(\mathbf{r}_{12}) \equiv \mathbf{r}_{12}\cdot\bp_{1}/(2\pi r_{12}^{2})$, obtained from Eqs.~\eqref{eq:multip-2} and \eqref{eq:Pphi}, denotes the change in the local pressure when only body~$1$ is present in the fluid. Using the definition ${\bf F}_{12}\equiv\bp_{2}-\tilde{\bp}_{2}$ in Eq.~\eqref{eq:pexp},
we arrive at 
\begin{align}
{\bf F}_{12} & =\mathbf{R}_{2}^{P}\Delta P_{1}\!\left(\br_{12}\right)-\mathbb{R}_{2}\frac{\mu}{2\pi}\left[\frac{\bp_{1}}{r_{12}^{2}}-\frac{2\left(\mathbf{r}_{12}\cdot\bp_{1}\right)\mathbf{r}_{12}}{r_{12}^{4}}\right]+\mo{r_{12}^{-2}}.
\end{align}
The force $\mathbf{F}_{12}$ can now be decomposed according to Eq.~\eqref{eq:dec}, in which $\mathbf{F}_{12}^{a}$ acts solely on asymmetric
bodies ($\mathbf{p}_{2}\neq0$) and $\mathbf{F}_{12}^{s}$ acts even
on fully symmetric bodies ($\bp_{2}=0$). We find these to be given
by 
\begin{align}
\mathbf{F}_{12}^{a} & =\mathbf{R}_{2}^{P}\Delta P_{1}\!\left(\br_{12}\right)+\mathcal{O}\left(r_{12}^{-2}\right)\label{eq:Fa}\\
\mathbf{F}_{12}^{s} & =-\mathbb{R}_{2}\frac{\mu}{2\pi}\left[\frac{\bp_{1}}{r_{12}^{2}}-\frac{2\left(\mathbf{r}_{12}\cdot\bp_{1}\right)\mathbf{r}_{12}}{r_{12}^{4}}\right]+\mathcal{O}\left(r_{12}^{-3}\right),\label{eq:Fs}
\end{align}
Using the single-body result Eq.~\eqref{eq:Jff}, the second equality can also be written as $\mathbf{F}_{12}^{s}=\mathbb{R}_{2}\bj_{1}\!\left(\br_{12}\right)+\mathcal{O}\!\left(r_{12}^{-3}\right)$, where $\bj_{1}$ denotes the current field induced by body~$1$ alone. Thus we have finally derived Eqs.~\eqref{eq:Fa-1} and~\eqref{eq:Fs-1}.

As noted before, Eq.~\eqref{eq:Fa} can be rewritten in terms of a linear response to the density modulation.
Under the assumption that the fluid has only a single homogeneous phase,
$P\!\left(\rho_{b}\right)$ is bound to be a strictly monotonically increasing function of $\rho_{b}$. Thus
$P\!\left(\rho_{b}\right)$ is invertible, allowing us to rewrite Eq.~\eqref{eq:response} as
\begin{align}
\mathbf{R}_{j}^{P} & =\frac{1}{P'\!\left(\rho_{b}\right)}\int d^{2}\br\left.\partial_{\rho_{b}}\rho\left[\rho_{b},\bj_{b}\right]\right|_{\bj_{b}=0}\bnabla V_{j}.
\end{align}
Combining this with Eqs.~\eqref{eq:ffrho} and \eqref{eq:Rjrho} yields Eq.~\eqref{eq:responserho}, which is the density version of Eq.~\eqref{eq:Fa}.

\subsection{Torque}

To obtain the interaction torques mediated by the active particles, we need to derive an expression for the density shift near one body, say body~$2$, induced by the presence of the other body, say body $1$. Substituting Eq.~\eqref{eq:Pdect} in Eq.~\eqref{eq:Pphi}, we get
\begin{align}
P & =P\!\left(\rho_{b}\right)+\Delta\tilde{\Phi}_{1}+\mo{\Delta\tilde{\Phi}_{1}^{2},\partial\Delta\tilde{\Phi}_{1},\Delta\tilde{\Phi}_{2}}.
\end{align}
Inserting this into the inverted expansion Eq.~\eqref{eq:rhoP}, we obtain
\begin{align}
\tilde{\rho} & =\rho_{b}+\frac{\Delta\tilde{\Phi}_{1}}{P'\!\left(\rho_{b}\right)}+\mo{\Delta\tilde{\Phi}_{1}^{2},\partial\Delta\tilde{\Phi}_{1},\Delta\tilde{\Phi}_{2}},\label{eq:rhoP1P2}
\end{align}
where a tilde above $\rho$ indicates that this is a solution of the two-body problem. Meanwhile, using Eqs.~\eqref{eq:dPb-2} and~\eqref{eq:dJb2} in Eq.~\eqref{eq:multip-1-1-1} gives
\begin{align}
\Delta\tilde{\Phi}_{1} & =\Delta P_{1}\!\left(\br_{12}\right)+\frac{\bj_{1}\!\left(\br_{12}\right)\cdot\br}{\mu}+\mo{r_{12}^{-2},r_{12}^{-3}r^{2}}.
\end{align}
Using this relation in Eq.~\eqref{eq:rhoP1P2}, we can write
\begin{align}
\tilde{\rho} & =\rho_{b}+\frac{\Delta P_{1}\!\left(\br_{12}\right)}{P'\!\left(\rho_{b}\right)}+\frac{\bj_{1}\!\left(\br_{12}\right)\cdot\br}{\mu P'\!\left(\rho_{b}\right)}+\mo{r_{12}^{-2},r_{12}^{-3}r^{2},r^{-1}},\label{eq:rhoexp-1}
\end{align}
where the scaling of the higher-order corrections can be justified by the multipole expansion of $\Delta\tilde{\Phi}_{1}$ shown in Eq.~\eqref{eq:multip-1-1-1} and the corresponding expansion of $\Delta\tilde{\Phi}_{2}$ that can be obtained by exchanging the indices $1$ and $2$. Using the single-body result Eq.~\eqref{eq:ffrho} for body~$1$, we can also write 
\begin{align}
\tilde{\rho} & =\rho_{b}+\delta\rho_{1}\!\left(\br_{12}\right)+\frac{\bj_{1}\!\left(\br_{12}\right)\cdot\br}{\mu P'\!\left(\rho_{b}\right)}+\mo{r_{12}^{-2},r_{12}^{-3}r^{2},r^{-1}}.\label{eq:rhoexp-1-1}
\end{align}
Therefore, to leading order in $r_{12}$, $\tilde{\rho}$ has modified
boundary conditions associated with a local density shift
$\delta\rho_{1}\!\left(\br_{12}\right)$ and a local current
$\bj_{1}\!\left(\br_{12}\right)$. An expansion with respect to
these changes gives
\begin{align}
\tilde{\rho} & =\left[1+\Delta P_{1}\!\left(\br_{12}\right)\partial_{P\!\left(\rho_{b}\right)}+\bj_{1}\!\left(\br_{12}\right)\cdot\bnabla_{\bj_{b}}\right]\left.\rho\!\left[P\!\left(\rho_{b}\right),\bj_{b}\right]\right|_{\bj_{b}=0}+\mo{r_{12}^{-2}}\label{eq:Prho}\\
 & =\left[1+\delta\rho_{1}\!\left(\br_{12}\right)\partial_{\rho_{b}}+\bj_{1}\!\left(\br_{12}\right)\cdot\bnabla_{\bj_{b}}\right]\left.\rho\!\left[\rho_{b},\bj_{b}\right]\right|_{\bj_{b}=0}+\mo{r_{12}^{-2}},\label{eq:rhorho}
\end{align}
where we have used Eq.~\eqref{eq:rhoexp-1} to obtain the
first equality and Eq.~\eqref{eq:rhoexp-1-1} to derive the
second.

We can now use Eqs.~\eqref{eq:Prho} or
\eqref{eq:rhorho} to find the interaction torque $\bm{\tau}_{12}$ applied
by body $1$ on body $2$. The self-torque experienced by body $j$
in the two-body problem is
\begin{align}
\tilde{\bm{\tau}}_{j} & =\int d^{2}\mathbf{r}\,\tilde{\rho}\!\left(\mathbf{r}\right)\left(\mathbf{r}-{\bf X}_{j}\right)\times\bm{\nabla}V_{j}\left(\br\right).\label{eq:taut}
\end{align}
Using Eqs.~\eqref{eq:Prho} and \eqref{eq:rhorho} in the above equation, we then obtain
\begin{align}
\bm{\tau}_{12} &\equiv \bm{\tau}_{2}-\tilde{\bm{\tau}}_{2}\nonumber \\
 &=\mathbf{T}_{2}^{P}\,\Delta P_{1}\!\left(\br_{12}\right)+\bm{\gamma}_{2}\times\bj_{1}\left(\br_{12}\right)+\mathcal{O}\!\left(r_{12}^{-2}\right)\\
 &=\mathbf{T}_{2}^{\rho}\,\delta \rho_{1}\!\left(\br_{12}\right)+\bm{\gamma}_{2}\times\bj_{1}\left(\br_{12}\right)+\mathcal{O}\!\left(r_{12}^{-2}\right),
\end{align}
where $\mathbf{T}_{2}^{P}$, $\bm\gamma_2$, and $\mathbf{T}_{2}^{\rho}$ are as defined in Eqs.~\eqref{eq:response-2}, \eqref{eq:gamma}, and \eqref{eq:trho}, respectively. As was the case for $\mathbf{F}_{12}$, $\bm{\tau}_{12}$ can also be decomposed into two components shown in Eqs.~\eqref{eq:tdec}, \eqref{eq:ta}, and \eqref{eq:ts}, so that $\bm{\tau}_{12}^{a}$ acts solely on bodies with a nonzero self-torque ($\bm{\tau}_{2}\neq0$), whereas $\bm{\tau}_{12}^{s}$ acts even on bodies with no self-torque ($\bm{\tau}_{2}=0$).

\section{Conclusions}
\label{sec:conclusions}

In this paper, we have studied the long-range effects of passive bodies immersed in a fluid of mutually interacting active particles. We have shown that, to leading order in an asymptotic far-field expansion, an asymmetric body generates dipolar density and pressure gradients as well as currents, all of which decay as a power law with increasing distance. These fields mediate generic long-range interactions between the passive bodies, which also decay algebraically with distance and do not obey an action--reaction principle. Remarkably, the leading-order behaviors of these interactions can be predicted by numerically or empirically measuring a few single-body properties in separate experiments. Our results provide a natural extension of the previous results obtained for ideal active fluids~\cite{BaekPRL2018}. While the interparticle interactions do not alter the symmetry and scaling exponents of the leading-order behaviors, they do modify the amplitudes of the long-range effects via nonideal behaviors of pressure. We recall that the interactions mediated by ideal active fluids induce interesting dynamical effects~\cite{BaekPRL2018} with possible applications to the flocking of shaken granular media~\cite{KumarNatComm2014} and the control of self-assembly by tuning the body shapes~\cite{AransonCRP2013,SotoPRL2014,SotoPRE2015,YangSM2014,MalloryARPC2018,GlotzerSc2004}. Our results clarify how such effects can be enhanced or inhibited by choosing the interparticle interactions of the active fluid. It will be very interesting to observe long-range currents and density modulations in experiment, an effort which could lead toward the useful applications described above.

Notably, our derivations of the leading-order long-range interactions rely solely on the assumption that the active fluid is deep inside the disordered phase, is far from the critical point (if any), and has a stress expansion shown in Eq.~\eqref{eq:sigasum}. Any overdamped system capable of demonstrating ratchet-like effects satisfying these assumptions exhibits the same phenomena, irrespective of the details of its constitutive relations. This is the case even if the interparticle interaction is dependent on the positions of arbitrarily many particles ({\it i.e.}, it is not a pairwise interaction) as long as it has a short range.

This study can still be extended in various directions. For example, it should be noted that the derivation does not work for interactions involving internal degrees of freedom, such as quorum sensing~\cite{MillerARM2001,LiuSCI2011,ThompsonJSM2011,FuPRL2012,CatesRPP2012,FarrellPRL2012,SotoPRE2014,SolonEPJ2015,SolonNJP2018},   orientational alignment~\cite{ViceskPRL1995,PeruaniEPJ2008,ChepizhkoPA2010,LiebchenPRL2017,LevisJPCM2018,MartinSM2018} and nematic alignment~\cite{RamaswamyEPL2003,HatwalnePRL2004,BertinNJP2013,BechingerRMP2016,DoostmohammadiNATCOM2018}. It will be interesting to check if active fluids with such interparticle interactions, especially those with symmetry-breaking transitions producing orientational order, can mediate novel kinds of long-range forces and torques. Further, one can examine the consequences of introducing the bodies into a critical or super-critical fluid undergoing MIPS, a subject of various recent theoretical advancement \cite{CatesARCMP2015,SolonPRE2018,SolonNJP2018,DigregorioPRL2018}. One can also consider long-range interparticle interactions, such as hydrodynamic interactions, instead of the short-range interactions considered here. This can be applicable to active particles suspended in a momentum-conserving, \textit{wet}, passive bath~\cite{BechingerRMP2016,BrottoPRL2013,HennesPRL2014,MatasPRE2014,ElgetiRPP2015,TiribocchiPRE2015,YoshinagaPRE2017,SaintillanARFM2018,WangCM2019}. We also note that, unlike the leading-order components of the interactions, higher-order terms may exhibit features which are qualitatively different from the non-interacting case. In fact, previous numerical and experimental studies show that near-field interactions can be attractive, both inherently and due to depletion forces~\cite{BechingerRMP2016,AngelaniPRL2011,ParraPRE2014,RayPRE2014,HarderJCP2014,RanPRL2015,LeitePRE2016}. That said, considering such higher-order far-field effects and near-field effects should unveil even richer physics of bodies immersed in active fluids.

\noindent {\it Acknowledgments:}
We thank Tal Agranov for the critical reading of the paper. O.G. and Y.K. are supported by an NSF-BSF grant and an ISF grant. Y.B. acknowledges that this work was supported by the New Faculty Startup Fund from Seoul National University. Y.B. is also supported in part by the European Research Council under the Horizon 2020 Program, ERC Grant Agreement No. 740269.

\appendix

\part*{Appendix}

\section{Weak-interaction expansions}
\label{sec:stress}

Here we present two explicit mathematical justifications of Eq.~\eqref{eq:sigasum}. The first one is obtained by taking the limit of weak pairwise interactions between active particles. This is a standard procedure, well studied in the context of equilibrium systems. We remind the reader that the important justification is that the verification of the results through the numerics, which extends outside of the weak-interactions regime. Revisiting Eq.~\eqref{eq:deanmn-2-1} and using integration by parts
\begin{align}
-\int d^{2}\br'\left[\bnabla U\!\left(|\br-\br'|\right)\right]\braket{\hat{\mb}^{\left(n\right)}\left(\br\right)\hat{\rho}\!\left(\br'\right)} & =- \int d^{2}\br'\,U\!\left(|\br-\br'|\right)\bnabla'\braket{\hat{\mb}^{\left(n\right)}\left(\br\right)\hat{\rho}\!\left(\br'\right)}.
\end{align}
For short-ranged $U$ we can interpret Eq.~\eqref{eq:deanmn-2-1} as a recurrence relation~\cite{NikolaPRL2016,BaekPRL2018}
\begin{align}
\mb^{\left(n\right)} & =\mb^{\left(n\right)}\!\left(\partial\mb^{\left(n-1\right)},\partial\mb^{\left(n+1\right)},\partial^{2}\mb^{\left(n\right)},\partial^{2}\braket{\hat{\mb}^{\left(n\right)}\hat{\rho}},\partial^{3}\braket{\hat{\mb}^{\left(n\right)}\hat{\rho}},\ldots\right)\label{eq:rec}
\end{align}
for $n \ge 1$. We note that the equation for $n = 0$ is set by Eq.~\eqref{eq:cont}. By unfolding Eq.~\eqref{eq:rec}, one can also write for $n \ge 1$
\begin{align}
\mb^{\left(n\right)} & =\mb^{\left(n\right)}\!\left(\partial^{n}\rho,\,\partial^{n+1}\rho,\dots;\text{ pair correlations and their derivatives}\right).
\label{eq:hierarchy}
\end{align}
Using this relation in Eqs.~\eqref{eq:sigma}, \eqref{eq:sigmap}, and \eqref{eq:IK}, the dependence of $\bm{\sigma}$ on the field variables can be written as
\begin{align}
\bm{\sigma} & =\bm{\sigma}\!\left(\rho,\partial\rho,\partial^{2}\rho,\ldots;\text{ pair correlations and their derivatives}\right).\label{eq:sigdep}
\end{align}
In the far-field, where one expects the deviations from the homogeneous density $\rho_{b}$ to be small, a standard dimensional analysis yields $\braket{\hat{\mb}^{\left(n\right)}\!\left(\br\right)\hat{\rho}\!\left(\br'\right)}=\mo{\rho_{b}^{2}}$ almost everywhere. As is evident from the forms of Eqs.~\eqref{eq:deanmn-2-1}, \eqref{eq:sigma}, \eqref{eq:sigmap}, and \eqref{eq:IK}, contributions from pair correlations always involve a factor of $U_{0}\equiv\int d^{2}\br\, U\!\left(\br\right)$, which has dimension of energy times area and is finite if $U$ has a short range. Thus, taking the weak-interaction limit amounts to assuming
that the dimensionless parameter $U_{0}\rho_{b}/T_\text{eff}$ is small. Note that the zero-order expansion trivially corresponds to the non-interacting limit.

Our derivation of Eq.~\eqref{eq:sigasum} from Eq.~\eqref{eq:sigdep} is described as follows. In \ref{sec:wie}, we show
how rapidly two-point correlations decay with increasing distances
from the body and decreasing magnitude of
interparticle interactions, namely
\begin{align} \label{eq:corr_decay}
\braket{\hat{\rho}\!\left(\br\right)\hat{\mb}^{(n)}\!\left(\br'\right)}_c \equiv \braket{\hat{\rho}\!\left(\br\right)\hat{\mb}^{(n)}\!\left(\br'\right)}-\rho\!\left(\br\right)\mb^{(n)}\!\left(\br'\right)=\mathcal{O}\!\left(\min\!\left(r^{-3},r'^{-3}\right),U_{0}\rho_{b}^{3}\right)
\end{align}
for any nonnegative integer $n$. We achieve this by deriving
the dynamics of the two-point correlations $\braket{\hat{\mb}^{\left(n\right)}\!\left(\br\right)\hat{\rho}\left(\br'\right)}$ and neglecting terms which are $\mathcal{O}\left[\left(U_{0}\rho_{b} / T_\text{eff}\right)^{2}\right]$.
Extension to higher orders can be done similarly by constructing a dynamical
BBGKY hierarchy of correlations. Using our result Eq.~\eqref{eq:corr_decay}, we show that two-point correlations yield
only subleading contributions to the stress tensor. Namely, we show that Eq.~\eqref{eq:sigdep} reduces to
\begin{align}
\frac{1}{T_\text{eff}\,\rho_{b}}\bm{\sigma} =\frac{1}{T_\text{eff}\,\rho_{b}}\bm{\sigma}\!\left(\rho\right)+\mathcal{O}\left[\partial\rho,\left(\frac{U_{0}\rho_{b}}{T_\text{eff}}\right)^2\right],\label{eq:wiexp}
\end{align}
which reproduces Eq.~\eqref{eq:sigasum}. Here, $\bm{\sigma}(\rho)$ denotes $\bm{\sigma}(\rho,0,0,\ldots)$. In \ref{sec:derv-vir}, we show that the first-order contribution of $U_{0}\rho_{b}/T_\text{eff}$, already
contained within $\bm{\sigma}\!\left(\rho\right)$, changes the pressure according to
\begin{align}
\frac{P\!\left(\rho_{b}\right)}{T_\text{eff}\,\rho_{b}} &= 1 +\frac{U_{0}\rho_{b}}{2T_\text{eff}}+\mathcal{O}\!\left[\left(\frac{U_{0}\rho_{b}}{T_\text{eff}}\right)^{2}\right],\label{eq:vdw}
\end{align}
which has the form of a standard virial expansion of
a van der Waals gas at temperature $T_\text{eff}$.

Instead of the weak interactions limit, we can also use a mean-field
approximation to write $\braket{\hat{\mb}^{\left(n\right)}\left(\br\right)\hat{\rho}\!\left(\br'\right)}\approx\mb^{\left(n\right)}\!\left(\br\right)\rho\!\left(\br'\right)$.
In this case the stress tensor in Eq.~\eqref{eq:sigdep} does not depend on pair correlations and can be expressed as
\begin{align}
\bm{\sigma} & \approx\bm{\sigma}\!\left(\rho,\partial\rho,\partial^{2}\rho,\ldots\right),
\end{align}
reproducing Eq.~\eqref{eq:sigasum} once more. This also yields the van der Waals equation~\eqref{eq:vdw}, except that the mean-field approach does not require a small dimensionless parameter.

\subsection{Weak-interaction expansion of the stress tensor}\label{sec:wie}

To derive Eq.~\eqref{eq:corr_decay}, we need to examine
the dynamics of $\braket{\hat{\rho}\!\left(\br\right)\hat{\rho}\!\left(\br'\right)}$ and impose the steady-state constraint.
For simplicity, we consider the case of ABPs ($\alpha=0$),
so that one can make use of standard It\^{o} calculus of continuous processes,
as previously demonstrated for passive particles~\cite{DeanJPA1996}.
Our key result is the dipolar decay of $\braket{\hat{\rho}\!\left(\br\right)\hat{\mb}^{(n)}\left(\br'\right)}_c$ with
increasing distances from the origin in the four-dimensional space
$\left(\br,\br'\right)$, which is much faster than the corresponding
decay in two dimensions. We note that similar results were also
obtained in other diffusive systems~\cite{SadhuPRE2014,BodineauJSP2010}.

As a first step, we examine the time evolution
of the empirical distribution of particles at position $\br$ and
orientation $\theta$, $\hat{\psi}\!\left(\mathbf{r},\theta\right)\equiv\sum_{i} \delta\!\left(\br-\br_{i}\right)\delta\!\left(\theta-\theta_{i}\right)$.
Through a standard procedure based on It\^o calculus,
as explicitly formulated by Dean~\cite{DeanJPA1996}
(see also \cite{SolonPRL2015,NikolaPRL2016}),
the time evolution of $\hat{\psi}$ is derived from
Eqs.~\eqref{eq:langevinr} and \eqref{eq:langevint} as
\begin{align}
\partial_{t}\hat{\psi} & =-\bm{\nabla}\cdot\left[ v\mathbf{e}_{\theta}-\mu\bm{\nabla}V-\mu\int d^{2}\mathbf{r}'\int d\theta'\,\hat{\psi}\!\left(\mathbf{r}',\theta'\right)\bnabla U\!\left(\left|\mathbf{r}-\mathbf{r}'\right|\right)-D_\text{t}\bm{\nabla}\right] \hat{\psi}(\br,\theta)\nonumber\\
&\quad +\bm{\nabla}\cdot\sqrt{2D_\text{t}\hat{\psi}}\,\hat{\bm{\eta}}+\partial_{\theta}\left(D_\text{r}\partial_{\theta}\hat{\psi}+\sqrt{2D_\text{r}\hat{\psi}}\,\hat{\xi}\right),\label{eq:deans}
\end{align}
where $\hat{\bm{\eta}}$ and $\hat{\xi}$ are Gaussian white noise fields
with unit amplitude. It should be noted that,
if one strictly carries out the derivation, $\hat{\psi}\!\left(\br,\theta\right)\hat{\psi}\!\left(\br',\theta'\right)$
in the above expression should be replaced with
$\hat{\psi}\!\left(\br,\theta\right)\hat{\psi}\!\left(\br',\theta'\right)-\psi\!\left(\br,\theta\right)\delta\!\left(\br-\br'\right)/2\pi$.
The extra term reflects the fact that a particle cannot exert
a force on itself. For simplicity, we eliminate this correction
by assuming $\bnabla U\!\left(0\right) = \mathbf{0}$,
which is naturally true for a smooth, spherically symmetric
interaction potential.

The empirical distribution $\hat{\psi}$ can be decomposed
into the Fourier components
\begin{align}
\int d\theta\,\hat{\psi}\!\left(\mathbf{r},\theta\right)\mathbf{e}_{n\theta} = \hat{\mathbf{m}}^{\left(n\right)}\!\left(\mathbf{r}\right),\label{eq:mnn}
\end{align}
with $\mb^{\left(n\right)}$ being the marginal empirical distributions defined in Eq.~\eqref{eq:marg}. In particular, for $n=0$ and $1$, these are related to the empirical density and empirical polarization density by
\begin{align}
\hat{\rho}\!\left(\mathbf{r}\right)= \int d\theta\,\hat{\psi}\!\left(\mathbf{r},\theta\right), \quad
\hat{\mb}\!\left(\br\right) = \int d\theta\,\hat{\psi}\!\left(\mathbf{r},\theta\right)\mathbf{e}_{\theta},
\end{align}
which correspond to the empirical density and polarization fields,
respectively. Multiplying Eq.~\eqref{eq:deans} side by side with
$\mathbf{e}_{n\theta}$ and integrating over $\theta$, one obtains
the equations governing the time evolution of $\hat{\mb}^{\left(n\right)}$.

For $n=0$, we obtain a noisy continuity equation
\begin{align}
\partial_{t}\hat{\rho} & +\bm{\nabla}\cdot\hat{\mathbf{J}}=0,\label{eq:cont-1}
\end{align}
where the fluctuating current field is given by
\begin{align}
\hat{\mathbf{J}}\!\left(\mathbf{r}\right) & \equiv v\hat{\mathbf{m}}\!\left(\mathbf{r}\right)-\mu\hat{\rho}\!\left(\mathbf{r}\right)\bm{\nabla}\left[V\!\left(\mathbf{r}\right)+\int d^{2}\mathbf{r}'\hat{\rho}\!\left(\mathbf{r}'\right)U\!\left(\left|\mathbf{r}-\mathbf{r}'\right|\right)\right]-D_\text{t}\bm{\nabla}\hat{\rho}\!\left(\mathbf{r}\right)+\sqrt{2D_\text{t}\hat{\rho}\left(\mathbf{r}\right)}\,\hat{\bm{\chi}}\!\left(\mathbf{r},t\right),\label{eq:current-1}
\end{align}
with the Gaussian white noise field $\sqrt{\hat{\rho}\!\left(\mathbf{r}\right)}\,\hat{\bm{\chi}}\!\left(\mathbf{r},t\right)\equiv\int d\theta\,\sqrt{\hat{\psi}\!\left(\br,\theta\right)}\,\hat{\bm{\eta}}\!\left(\br,t\right)$ satisfying
\begin{align}
\left\langle \bm{\nabla}\cdot\sqrt{\hat{\rho}\!\left(\mathbf{r}\right)}\,\bm{\chi}\!\left(\mathbf{r},t\right)\bm{\nabla}'\cdot\sqrt{\hat{\rho}\!\left(\mathbf{r}'\right)}\,\bm{\chi}\!\left(\mathbf{r}',t'\right)\right\rangle  & =\frac{1}{2}\id\left[-\nabla_{\left(\br,\br'\right)}^{2}\delta\!\left(\mathbf{r}-\mathbf{r}'\right)\rho\!\left(\br\right)+\delta\!\left(\mathbf{r}-\mathbf{r}'\right)\lap\rho\!\left(\br\right)\right]\delta\!\left(t-t'\right).\label{eq:chicor}
\end{align}
From here on, we define and use a four-dimensional differential operator
$\bnabla_{\left(\br,\br'\right)} \equiv \bnabla\oplus\bnabla' = (\bnabla,\bnabla')^T$,
which also implies $\nabla_{\left(\br,\br'\right)}^{2}=\lap+\nabla'^{2}$.

Similarly, for $n=1$, we obtain
\begin{align}
l_\text{r}\partial_{t}\hat{\mathbf{m}} & =-v\hat{\mathbf{m}}+\mu l_\text{r}\bnabla\cdot\left[(\bnabla V)\hat{\mathbf{m}}\right]+\mu\bnabla\cdot\hat{\bm\sigma}^\text{P}-\frac{vl_\text{r}}{2}\bnabla\hat{\rho}+l_\text{r}\sqrt{2D_\text{r}}\,\hat{\bm\xi}^{(1)},\label{eq:vm-1}
\end{align}
where $\hat{\bm{\xi}}^{(1)}(\br,t)\equiv\int d\theta\,\sqrt{\hat{\psi}\!\left(\br,\theta\right)}\,\hat{\xi}\!\left(\br,\theta,t\right)\bf{e}^{\perp}_\theta$ with $\hat{\bf{e}}^{\perp}_\theta\equiv\mathbf{e}_{z}\times\hat{\bf{e}}_\theta=\left(-\sin\theta,\cos\theta\right)^T$,
and we have defined the polarization component of
the stress tensor (the noisy counterpart of Eq.~\eqref{eq:sigmap})
\begin{align}
\hat{\bm\sigma}^\text{P}(\br) & \equiv l_\text{r}\int d^{2}\mathbf{r}'\,\left[\bnabla U\!\left(\left|\mathbf{r}-\mathbf{r}'\right|\right)\right]\hat{\mathbf{m}}\!\left(\br\right)\hat{\rho}\!\left(\mathbf{r}'\right)+Tl_\text{r}\bnabla\hat{\mathbf{m}}\!\left(\br\right)-2\left(T_\text{eff}-T\right)\hat{\mathbb{Q}}\!\left(\br\right)-\frac{l_\text{r}}{\mu}\sqrt{2D_\text{t}}\,\hat{\bm\chi}^{\left(1\right)}(\br,t),
\end{align}
with $\hat{\bm\chi}^{\left(1\right)}(\br,t)\equiv\int d\theta\,\sqrt{\hat{\psi}\!\left(\br,\theta\right)}\,\hat{\bm\eta}\!\left(\br,\theta,t\right)\bf{e}_\theta$.
Finally, the nematic order tensor $\hat{\mathbb{Q}}$
is again defined by Eq.~\eqref{eq:nem}, which can also be written
in terms of $\hat{\psi}$ as
\begin{align}
\hat{\mathbb{Q}}\!\left(\br\right)=\int d\theta\left(\mathbf{e}_{\theta}\mathbf{e}_{\theta}-\frac{1}{2}\id\right)\hat{\psi}\!\left(\mathbf{r},\theta\right).
\end{align}
Note that by taking the average of Eq.~\eqref{eq:vm-1} at steady-state, at which $\partial_t \braket{\hat{\bf{m}}}=\partial_t \bf{m}=0$, and combining with Eq.~\eqref{eq:current-1}, we recover Eqs.~\eqref{eq:cdyn-1} and~\eqref{eq:sigma} of Section \ref{sec:hydrodynamics}. Furthermore, one can reach Eqs.~\eqref{eq:deanmn-2-1} and \eqref{eq:vm} by using Eq.~\eqref{eq:mnn} for arbitrary $n$ and noting that $\bnabla\cdot\hat{\mathbb{Q}}=-\mathbb{D}^\dagger \hat{\mb}^{\left(2\right)}/2$.

\begin{figure}
  \centering
  \includegraphics[width=0.9\textwidth]{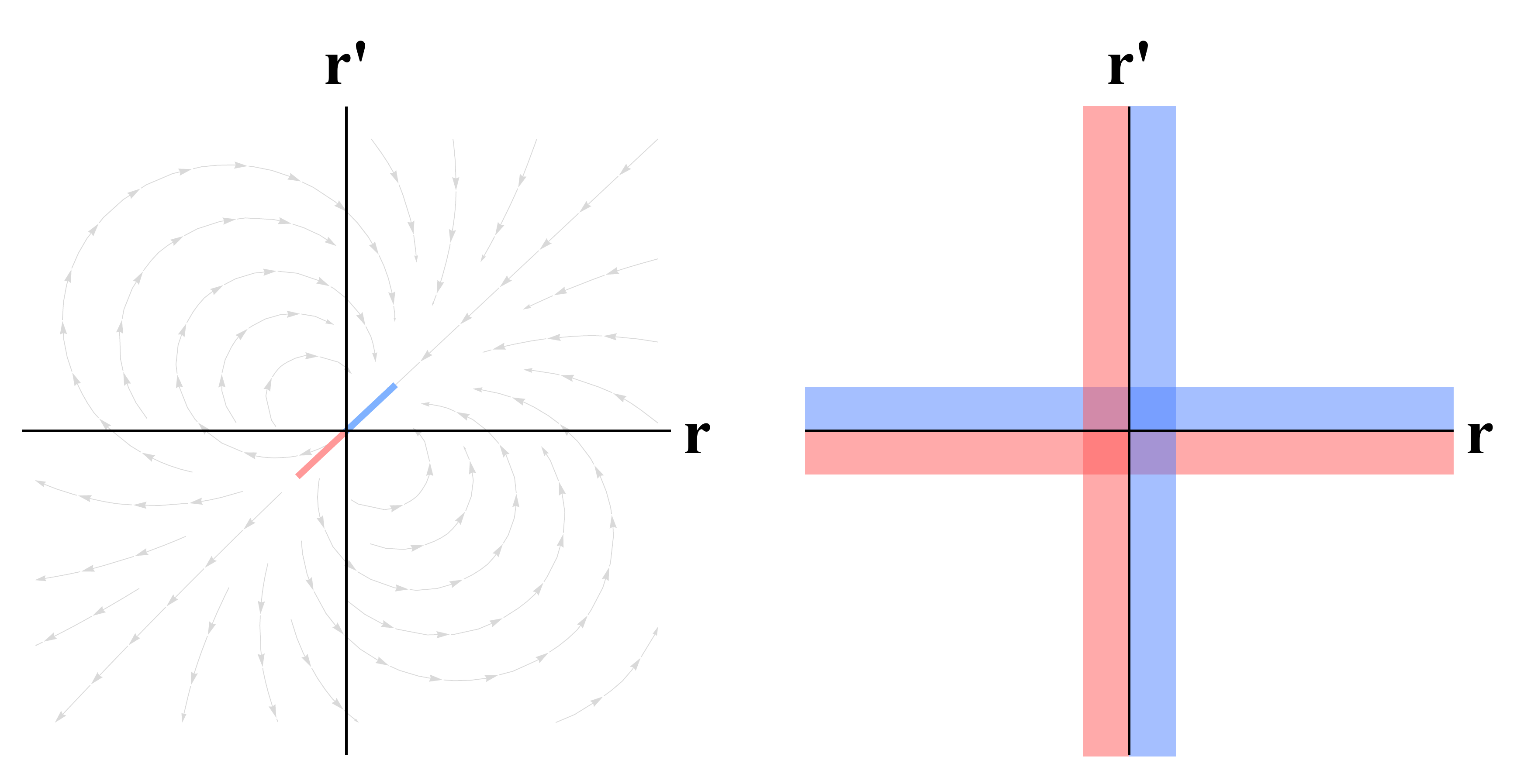}
  \caption{Schematic representation of the leading-order charge distribution in Eqs.~\eqref{eq:J2c-1}, \eqref{eq:Pdist} and \eqref{eq:Idist} (red and blue colors). Left: the charge density $\mathcal{I}\left(\br,\br'\right)$. A pictorial description of the four-dimensional current $\bj^{(2)} \left(\br,\br'\right)$ is shown in grey arrows (see Eq.~\eqref{eq:cflux}). The charges are concentrated on the plane $\br=\br'$. Within the first order in the expansion, Eq.~\eqref{eq:laprho} holds, implying that the distribution is asymmetric and of length $\sim d$ (see text). Right: the charge density $\mathcal{P}\left(\br,\br'\right)$. Each dipole sheet is of thickness $\sim d$.}
  \label{fig:4d}
\end{figure}

Given these results for the single-point observables, we now move
on to the time evolution of two-point observables. Applying
Eq.~\eqref{eq:chicor}, It\^o's product rule for the time derivative of 
$\langle\hat{\rho}\!\left(\mathbf{r}\right)\hat{\rho}\!\left(\mathbf{r}'\right)\rangle$ reads
\begin{align}
\partial_{t}\langle\hat{\rho}\!\left(\mathbf{r}\right)\hat{\rho}\!\left(\mathbf{r}'\right)\rangle & = \langle\hat{\rho}\!\left(\mathbf{r}'\right)\partial_{t}\hat{\rho}\!\left(\mathbf{r}\right)\rangle+\langle\hat{\rho}\!\left(\mathbf{r}\right)\partial_{t}\hat{\rho}\!\left(\mathbf{r}'\right)\rangle+D_\text{t}\left[-\nabla_{\left(\br,\br'\right)}^{2}\delta\!\left(\mathbf{r}-\mathbf{r}'\right)\rho\!\left(\br\right)+\delta\!\left(\mathbf{r}-\mathbf{r}'\right)\lap\rho\!\left(\br\right)\right].\label{eq:prodrule}
\end{align}
Using Eqs.~\eqref{eq:cont-1} and \eqref{eq:current-1} to calculate
the first two terms on the rhs and evaluating the averages over
histories, we obtain the steady-state condition
\begin{align}
D_\text{t}\delta\!\left(\mathbf{r}-\mathbf{r}'\right)\lap\rho\!\left(\br\right) & =\bnabla_{\left(\br,\br'\right)}\cdot\bj^{\left(2\right)}\!\left(\br,\br'\right),\label{eq:cflux}
\end{align}
where the four-dimensional current density $\bj^{\left(2\right)}$
is given by
\begin{align}
\bj^{\left(2\right)}\!\left(\br,\br'\right) &\equiv v\left[
\braket{\hat{\rho}\!\left(\br'\right)\hat{\mb}\!\left(\br\right)}
\oplus\braket{\hat{\rho}\!\left(\br\right)\hat{\mb}\!\left(\br'\right)}\right]
-\mu\braket{\hat{\rho}\!\left(\br\right)\hat{\rho}\!\left(\br'\right)}\bnabla_{\left(\br,\br'\right)}\left[V\!\left(\br\right)+V\!\left(\br'\right)\right]\nonumber \\
 &\quad -\mu\int d^{2}\br''\braket{\hat{\rho}\!\left(\br\right)\hat{\rho}\!\left(\br'\right)\hat{\rho}\!\left(\br''\right)}\bnabla_{\left(\br,\br'\right)}\left[U\!\left(\left|\br-\br''\right|\right)+U\!\left(\left|\br'-\br''\right|\right)\right]\nonumber \\
 &\quad -D_\text{t}\bnabla_{\left(\br,\br'\right)}\left[\braket{\hat{\rho}\!\left(\br\right)\hat{\rho}\!\left(\br'\right)}-\delta\!\left(\mathbf{r}-\mathbf{r}'\right)\rho\!\left(\br\right)\right],\label{eq:j2}
\end{align}
which depends on $\braket{\hat{\rho}\!\left(\br\right)\hat{\mb}\!\left(\br'\right)}$. Note that the current density $\bj^{\left(2\right)}\left(\br,\br'\right)$
associated with the two-point correlation $\braket{\hat{\rho}\!\left(\br\right)\hat{\rho}\!\left(\br'\right)}$
has an asymmetric source, determined by $\lap\rho$ (see Fig.~\ref{fig:4d}). To obtain the steady-state expression for
$\braket{\hat{\rho}\!\left(\br\right)\hat{\mb}\!\left(\br'\right)}$,
we first need to examine its time evolution.
Using the cross correlation
\begin{align}
\braket{\bm{\nabla}\cdot\sqrt{\hat{\rho}\!\left(\mathbf{r}\right)}\,\hat{\bm{\chi}}\!\left(\mathbf{r},t\right)\,\bnabla'\cdot\hat{\bm{\chi}}^{\left(1\right)}\!\left(\mathbf{r'},t'\right)} & =\frac{1}{2}\left[-\nabla_{\left(\br,\br'\right)}^{2}\delta\!\left(\mathbf{r}-\mathbf{r}'\right)\mathbf{m}\!\left(\br\right)+\delta\!\left(\mathbf{r}-\mathbf{r}'\right)\lap \mathbf{m}\!\left(\br\right)\right]\delta\!\left(t-t'\right),
\end{align}
It\^{o}'s product rule yields
\begin{align}
\partial_{t}\left\langle\hat{\rho}\!\left(\mathbf{r}\right)\hat{\mb}\!\left(\mathbf{r}'\right)\right\rangle &= \left\langle\hat{\rho}\!\left(\mathbf{r}\right)\partial_{t}\hat{\mb}\!\left(\mathbf{r}'\right)\right\rangle+\left\langle\hat{\mb}\!\left(\mathbf{r}'\right)\partial_{t}\hat{\rho}\!\left(\mathbf{r}\right)\right\rangle+D_\text{t}\left[-\nabla_{\left(\br,\br'\right)}^{2}\delta\!\left(\mathbf{r}-\mathbf{r}'\right)\mb\!\left(\br\right)+\delta\!\left(\mathbf{r}-\mathbf{r}'\right)\lap\mb\!\left(\br\right)\right].\label{eq:prodrule-1}
\end{align}
In the steady state, using Eq.~\eqref{eq:cont-1} to eliminate $\partial_t\hat\rho(\mathbf{r})$ on the rhs, we obtain 
 \begin{align}
\braket{\hat{\rho}\!\left(\mathbf{r}\right)\partial_{t}\hat{\mb}\!\left(\mathbf{r}'\right)} = \bnabla\cdot\braket{\hat{\bj}\!\left(\br\right)\hat{\mb}\!\left(\mathbf{r}'\right)}-D_\text{t}\left[-\nabla_{\left(\br,\br'\right)}^{2}\delta\!\left(\mathbf{r}-\mathbf{r}'\right)\mb\!\left(\br\right)+\delta\!\left(\mathbf{r}-\mathbf{r}'\right)\lap\mb\!\left(\br\right)\right].\label{eq:cort}
\end{align}
Meanwhile, solving Eq.~\eqref{eq:vm-1} for $v\hat{\mathbf{m}}$ and
using the result in Eq.~\eqref{eq:j2} to rewrite the first term
on its rhs, we get
\begin{align}
\bj^{\left(2\right)}\!\left(\br,\br'\right) &= -l_\text{r}\left[\braket{\hat{\rho}\!\left(\br'\right)\partial_t\hat{\mb}\!\left(\br\right)}\oplus\braket{\hat{\rho}\!\left(\br\right)\partial_t\hat{\mb}\!\left(\br'\right)}\right] + \mu l_\text{r}\left\{\bnabla\cdot\left[\bnabla V\!\left(\br\right)\right]\braket{\hat{\rho}\!\left(\br'\right)\hat{\mb}\!\left(\br\right)}\oplus\bnabla'\cdot\left[\bnabla' V\!\left(\br'\right)\right]\braket{\hat{\rho}\!\left(\br\right)\hat{\mb}\!\left(\br'\right)}\right\} \nonumber\\
&\quad + \mu \left[\bnabla\cdot\braket{\hat{\rho}\!\left(\br'\right)\hat{\bm\sigma}^\text{P}\!\left(\br\right)}\oplus\bnabla'\cdot\braket{\hat{\rho}\!\left(\br\right)\hat{\bm\sigma}^\text{P}\!\left(\br'\right)}\right] -\mu\braket{\hat{\rho}\!\left(\br\right)\hat{\rho}\!\left(\br'\right)}\bnabla_{\left(\br,\br'\right)}\left[V\!\left(\br\right)+V\!\left(\br'\right)\right]\nonumber\\
&\quad -\mu\int d^{2}\br''\braket{\hat{\rho}\!\left(\br\right)\hat{\rho}\!\left(\br'\right)\hat{\rho}\!\left(\br''\right)}\bnabla_{\left(\br,\br'\right)}\left[U\!\left(\left|\br-\br''\right|\right)+U\!\left(\left|\br'-\br''\right|\right)\right]\nonumber \\
 &\quad -\bnabla_{\left(\br,\br'\right)}\left[D_{\text{eff}}\braket{\hat{\rho}\!\left(\br\right)\hat{\rho}\!\left(\br'\right)}-D_\text{t}\delta\!\left(\mathbf{r}-\mathbf{r}'\right)\rho\!\left(\br\right)\right],\label{eq:j2-1}
\end{align}
where $D_{\text{eff}}\equiv D_\text{t}+vl_\text{r}/2=\mu T_\text{eff}$
is the effective diffusion constant of active particles.
Using Eq.~\eqref{eq:cort} to eliminate both $\braket{\hat{\rho}\!\left(\br'\right)\partial_{t}\hat{\mb}\!\left(\br\right)}$ and $\braket{\hat{\rho}\!\left(\br\right)\partial_{t}\hat{\mb}\!\left(\br'\right)}$, we find
\begin{align}
\bj^{\left(2\right)}\left(\br,\br'\right) & =-\mu\braket{\hat{\rho}\!\left(\br\right)\hat{\rho}\!\left(\br'\right)}\bnabla_{\left(\br,\br'\right)}\left[V\!\left(\br\right)+V\!\left(\br'\right)\right]+\mu l_\text{r}\bnabla_{\left(\br,\br'\right)}\cdot\left\{\left[\bnabla V\!\left(\br\right)\right]\braket{\hat{\rho}\!\left(\br'\right)\hat{\mb}\left(\br\right)}\oplus\left[\bnabla' V\!\left(\br'\right)\right]\braket{\hat{\rho}\!\left(\br\right)\hat{\mb}\left(\br'\right)}\right\}\nonumber \\
 &   + l_\text{r}D_\text{t}\delta\!\left(\mathbf{r}-\mathbf{r}'\right)\nabla^2_{\left(\br,\br'\right)}\left[\mb\!\left(\br'\right)\oplus\mb\!\left(\br\right)\right]+\mu\bnabla_{\left(\br,\br'\right)}\cdot\bm{\sigma}^{\left(2\right)}\left(\br,\br'\right),\label{eq:J2f}
\end{align}
where we introduce the four-dimensional stress tensor
\begin{align}
\bm{\sigma}^{\left(2\right)}\left(\br,\br'\right) & \equiv-\left[T_\text{eff}\braket{\hat{\rho}\!\left(\br\right)\hat{\rho}\!\left(\br'\right)}-T\delta\!\left(\mathbf{r}-\mathbf{r}'\right)\rho\!\left(\br\right)\right]\id+\bm{\sigma}^{P(2)}\left(\br,\br'\right)+\bm{\sigma}^{IK(2)}\left(\br,\br'\right).\label{eq:sig2}
\end{align}
Here, the polarization tensor $\bm{\sigma}^{P(2)}$ is given by
\begin{align}
\bm{\sigma}^{P(2)}\left(\br,\br'\right) & \equiv-l_\text{r}T\bnabla_{\left(\br,\br'\right)}\delta\!\left(\mathbf{r}-\mathbf{r}'\right)\left[\mb\!\left(\br'\right)\oplus\mb\!\left(\br\right)\right]-\frac{l_\text{r}}{\mu}\gamma_0\cdot\left[\braket{\hat{\bj}\left(\br\right)\hat{\mb}\left(\mathbf{r}'\right)}\oplus\braket{\hat{\bj}\left(\br'\right)\hat{\mb}\left(\mathbf{r}\right)}\right]\nonumber\\ & +\braket{\hat{\rho}\!\left(\br'\right)\hat{\bm{\sigma}}^{P}\left(\br\right)}\oplus\braket{\hat{\rho}\!\left(\br\right)\hat{\bm{\sigma}}^{P}\left(\br'\right)},\label{eq:sigP2}
\end{align}
where we account for the transposed ordering of the direct sum by inserting the tensor product of the exchange tensor (first Pauli matrix) $\varsigma_{x}$ and the two-dimensional identity tensor $\id_2$,
\begin{align}
\gamma_0 & = \varsigma_{x}\id_2 =\begin{pmatrix}0 & 0 & 1 & 0\\
0 & 0 & 0 & 1\\
1 & 0 & 0 & 0\\
0 & 1 & 0 & 0
\end{pmatrix},
\end{align}
which is also the zeroth Dirac matrix in the chiral basis.
The interaction tensor $\bm{\sigma}^{IK(2)}$ satisfies
\begin{align}
\bnabla_{\left(\br,\br'\right)}\cdot\bm{\sigma}^{IK(2)}\left(\br,\br'\right) & \equiv-\int d^{2}\br''\braket{\hat{\rho}\!\left(\br\right)\hat{\rho}\!\left(\br'\right)\hat{\rho}\!\left(\br''\right)}\bnabla_{\left(\br,\br'\right)}\left[U\!\left(\left|\br-\br''\right|\right)+U\!\left(\left|\br'-\br''\right|\right)\right].
\end{align}

Up to this point, all of our results are exact.
From here on, to implement the weak-interaction assumption,
we neglect all terms of order $U_{0}$, which means that all interaction force integrals, including $\bm{\sigma}^{IK(2)}$,
are neglected. Noting that the steady-state condition $\bnabla\cdot\bj\!\left(\br\right)=0$
implies $\bnabla_{\left(\br,\br'\right)}\cdot\left[\rho\!\left(\br'\right){\bf J}\!\left(\br\right)\oplus\rho\!\left(\br\right){\bf J}\!\left(\br'\right)\right]=0$,
we can rewrite Eq.~\eqref{eq:cflux} as
\begin{align}
D_\text{t}\delta\!\left(\mathbf{r}-\mathbf{r}'\right)\lap\rho\!\left(\br\right) & =\bnabla_{\left(\br,\br'\right)}\cdot\bj_{c}^{\left(2\right)}\!\left(\br,\br'\right),\label{eq:cflux-1}
\end{align}
where $\bj_{c}^{\left(2\right)}\!\left(\br,\br'\right)\equiv{\bf J}^{\left(2\right)}\!\left(\br,\br'\right)-\left[\rho\!\left(\br'\right){\bf J}\left(\br\right)\oplus\rho\!\left(\br\right){\bf J}\!\left(\br'\right)\right]$.
Because Eq.~\eqref{eq:J2f} is separable, we use separation of variables and Eq.~\eqref{eq:cdyn-1} to obtain
\begin{align}
\bj^{\left(2\right)}_c\left(\br,\br'\right) & =-\mu\braket{\hat{\rho}\!\left(\br\right)\hat{\rho}\!\left(\br'\right)}_c\bnabla_{\left(\br,\br'\right)}\left[V\!\left(\br\right)+V\!\left(\br'\right)\right]+\mu l_\text{r}\bnabla_{\left(\br,\br'\right)}\cdot\left\{\left[\bnabla V\!\left(\br\right)\right]\braket{\hat{\rho}\!\left(\br'\right)\hat{\mb}\left(\br\right)}_c\oplus\left[\bnabla' V\!\left(\br'\right)\right]\braket{\hat{\rho}\!\left(\br\right)\hat{\mb}\left(\br'\right)}_c\right\}\nonumber \\
 &   + l_\text{r}D_\text{t}\delta\!\left(\mathbf{r}-\mathbf{r}'\right)\nabla^2_{\left(\br,\br'\right)}\left[\mb\!\left(\br'\right)\oplus\mb\!\left(\br\right)\right]+\mu\bnabla_{\left(\br,\br'\right)}\cdot\bm{\sigma}^{\left(2\right)}_c\left(\br,\br'\right),\label{eq:J2c}
\end{align}
where $\bm{\sigma}_{c}^{\left(2\right)}$ is given by the expression
for $\bm{\sigma}^{\left(2\right)}$ upon replacing all second-order correlations
with second-order cumulants, with the exception of the second term on the rhs of Eq.~\eqref{eq:sigP2}. Since at this order there is no long-range stress, we can
obtain $\braket{\hat{\rho}\!\left(\br\right)\hat{\rho}\!\left(\br'\right)}_{c}$
directly. Taking the divergence of Eq.~\eqref{eq:J2c} and using Eqs.~\eqref{eq:sig2} and~\eqref{eq:cflux-1}, we obtain the Poisson equation
\begin{align}
\nabla_{\left(\br,\br'\right)}^{2}\left[D_{\text{eff}}\braket{\hat{\rho}\!\left(\br\right)\hat{\rho}\!\left(\br'\right)}_{c}-D_\text{t}\rho\!\left(\br\right)\delta\!\left(\mathbf{r}-\mathbf{r}'\right)\right] & =\mu\left[\mathcal{P}\left(\br,\br'\right)+\mathcal{I}\left(\br,\br'\right)\right]+\mo{U_{0}\rho_{b}^{3}},\label{eq:J2c-1}
\end{align}
where we introduce the stress charge densities (see Fig.\ref{fig:4d})
\begin{align}
\mathcal{P}\left(\br,\br'\right) & \equiv -\bnabla_{\left(\br,\br'\right)}\cdot\left\{ \braket{\hat{\rho}\!\left(\br\right)\hat{\rho}\!\left(\br'\right)}_{c}\bnabla_{\left(\br,\br'\right)}\left[V\!\left(\br\right)+V\!\left(\br'\right)\right]\right\} \nonumber \\
 & +l_\text{r}\bnabla_{\left(\br,\br'\right)}\bnabla_{\left(\br,\br'\right)}:\left\{\left[\bnabla V\!\left(\br\right)\right]\braket{\hat{\rho}\!\left(\br'\right)\hat{\mb}\left(\br\right)}_{c}\!\left(\br\right)\oplus\left[\bnabla'V\!\left(\br'\right)\right]\braket{\hat{\rho}\!\left(\br\right)\hat{\mb}\left(\br'\right)}_{c}\right\}+\bnabla_{\left(\br,\br'\right)}\bnabla_{\left(\br,\br'\right)}:\bm{\sigma}_{c}^{P\left(2\right)},\label{eq:Pdist} \\
\mathcal{I}\left(\br,\br'\right) & \equiv -T\delta\!\left(\mathbf{r}-\mathbf{r}'\right)\lap\rho\!\left(\br\right)-l_\text{r}T\bnabla_{\left(\br,\br'\right)}\cdot\delta\!\left(\mathbf{r}-\mathbf{r}'\right)\left[\nabla'^2\mb\!\left(\br'\right)\oplus\nabla^2\mb\!\left(\br\right)\right].\label{eq:Idist}
\end{align}
Due to the separable nature Eq.~\eqref{eq:J2c-1}, the charges are concentrated in three sheets (see Fig.~\ref{fig:4d}). We will treat each of the two types of charge distribution separately, and show that the resulting solution decays as dipole in four-dimensions.

First, we claim that the charge density due to It{\^o} terms, $\mathcal{I}\left(\br,\br'\right)$, is localized in space and provides leading-order dipolar contributions. At this order in the weak-interaction expansion, $\mb$ and $\rho$ in Eqs.~\eqref{eq:cflux-1} and \eqref{eq:J2c} are the solutions of the corresponding non-interacting
problem. In the non-interacting problem, the angular hierarchy Eq.~\eqref{eq:hierarchy} becomes
\begin{align}
\mb^{\left(n\right)} & =\mb^{\left(n\right)}\!\left(\partial^{n}\rho,\,\partial^{n+1}\rho,\dots\right).
\label{eq:hierarchy1}
\end{align}
This allows one to represent $\lap\rho$ and $\lap\mb$ as a sum of terms proportional to $V$ and its derivatives. This procedure is the key step in writing the previously obtained solution to the non-interacting problem to arbitrary high order\cite{BaekPRL2018}. Specifically, for $\lap\rho$ we have
\begin{align}
D_{\text{eff}}\lap\rho&=-\mu\bnabla\cdot\left[\rho\left(\br\right)\bnabla V\!\left(\br\right)\right]+\mo{\partial^2},
\label{eq:laprho}
\end{align}
where $\mo{\partial ^2}$ indicates terms which are at least of second differential order.
Thus, $\lap\rho$
and $\lap\mb$ are short-ranged with characteristic length $d$, as they vanish quickly outside of the body. Moreover, the leading-order contribution from these is dipolar, as seen in Eq.~\eqref{eq:laprho}. This shows that $\mathcal{I}\left(\br,\br'\right)$ is indeed a localized density of leading-order dipolar contribution.

Next, we note that the separable density $\mathcal{P}\left(\br,\br'\right)$ is concentrated within two perpendicular charge sheets of thickness $d$. To leading-order, each sheet is a dipole sheet. By means of numerical solution and a self-consistent argument, it was previously shown that a Poisson equation with infinite sheets of multipole densities proportional to the potential, as in the above, yields a solution whose asymptotic behavior is that of a localized multipole of the same order~\cite{SadhuPRE2014}. We note that one can also verify this result using a weak-forcing expansion, where the small dimensionless parameter is $\rho_b V_0/T_\text{eff}$, $V_0\equiv\int d^2\br V\left(\br\right)$. At order $V_0^0$, all charge densities are neglected and we obtain $\braket{\hat{\rho}\!\left(\br\right)\hat{\rho}\!\left(\br'\right)}_{c}=\rho\!\left(\br\right)\delta\left(\br-\br'\right)T/T_\text{eff}$. At order $V_0^1$, $\mathcal{P}\left(\br,\br'\right)$ is obtained from the solution to the zero-order expansion, which amounts to an ideal dipole at the origin. Likewise, $\mathcal{I}\left(\br,\br'\right)$ now includes the localized dipolar contribution shown in Eq.~\eqref{eq:laprho}. The resulting asymptotic decay is that of a four-dimensional dipole, namely $\sim \left(r^2+r'^2\right)^{-3/2} \sim \min\!\left(r^{-3},r'^{-3}\right)$. At order $V_0^2$, the charge density is obtained from the solution to the first order expansion, giving a charge density that decays as $\sim r^{-3}$ and $\sim r'^{-3}$ respectively along each sheet. Then, we invoke the argument given in Ref.~\cite{SadhuPRE2014}, saying that a multipole density that decays faster than $r^{-2}$ induces a potential whose asymptotic behavior is that of a localized multipole\footnote{In Ref.~\cite{SadhuPRE2014}; Appendix C, the argument was given for a quadrupole density. However, the proof can be generalized to any multipole density in a direct way.}. By induction, the dipolar decay holds up to arbitrary order in the perturbative expansion. We conclude that $\mathcal{P}\left(\br,\br'\right)$ acts as an effectively localized dipole.

In total, the far-field behavior of the solution to Eq.~\eqref{eq:J2c} is given by
\begin{align}
\braket{\hat{\rho}\!\left(\br\right)\hat{\rho}\!\left(\br'\right)}_{c} & =\frac{T}{T_\text{eff}}\rho\!\left(\br\right)\delta\!\left(\mathbf{r}-\mathbf{r}'\right)+\mathcal{O}\left(\min\!\left(r^{-3},r'^{-3}\right),U_{0}\rho_{b}^{3}\right).
\label{eq:cumulant}
\end{align} 
Because the correlator $\braket{\hat{\rho}\!\left(\br\right)\hat{\rho}\!\left(\br'\right)}$ appears only within the interaction force density $\braket{\hat{\rho}\!\left(\br\right)\hat{\rho}\!\left(\br'\right)}\bnabla U\left(\br-\br'\right)$, and due to our assumption that $\bnabla U\!\left(0\right)=0$, we can omit the first term in Eq.~\eqref{eq:cumulant}. One can skip this
simplifying assumption if the above derivation is done for pair densities,
e.g. $\braket{\hat{\rho}\!\left(\br\right)\hat{\rho}\!\left(\br'\right)}-\rho\!\left(\br\right)\delta\!\left(\br-\br'\right)$,
instead of correlations. We conclude that $\braket{\hat{\rho}\!\left(\br\right)\hat{\rho}\!\left(\br'\right)}=\rho\!\left(\br\right)\rho\!\left(\br'\right)+\mo{\min\!\left(r^{-3},r'^{-3}\right),U_{0}\rho_{b}^{3}}$.

Similarly, one can derive hierarchical relations for two-point correlations,
as done for single-point averages in the above,
to obtain that $\braket{\hat{\rho}\!\left(\br\right)\hat{\mb}^{\left(n\right)}\left(\br'\right)}=\rho\!\left(\br\right)\mb^{\left(n\right)}\!\left(\br'\right)+\mo{\min\!\left(r^{-3},r'^{-3}\right),U_{0}\rho_{b}^{3}}$.
Lastly, we can utilize the fact that these correlators appear only within the interaction force densities $\braket{\hat{\rho}\!\left(\br\right)\hat{\mb}^{\left(n\right)}\left(\br'\right)}\bnabla U\left(\br-\br'\right)$ and that $U$ is short-ranged to replace the above corrections with $\mo{r^{-3},U_{0}\rho_{b}^{3}}$. This thereby confirms the stress expansion Eq.~\eqref{eq:wiexp} up to $\mo{r^{-3},\partial\rho,U_{0}^{2}\rho_{b}^{2}/T_\text{eff}^{2}}$. Since the correction is consistent with the rest of the derivation in the main text, it holds that $\rho-\rho_b\sim r^{-1}$.  It follows that the correction to the stress expansion is $\mo{\partial \rho,U_{0}^{2}\rho_{b}^{2}/T_\text{eff}^{2}}$, as written in Eq.~\eqref{eq:wiexp}.

The weak-interaction expansion can be extended into higher orders in the following way. Starting from Eq.~\eqref{eq:sig2}, one can repeat the process depicted in Section
\ref{sec:far-field} to show that the pressure field
$P_{c}^{\left(2\right)}\equiv-\tr\bm{\sigma}_{c}^{\left(2\right)}/4$
is given by
\begin{align}
P_{c}^{\left(2\right)}\!\left(\br,\br'\right) & =\frac{1}{4\pi^{2}}\int\frac{d^{2}\mathbf{s}\,d^{2}\mathbf{s}'}{\left|\mathbf{r}-\mathbf{s}\right|^{2}+\left|\mathbf{r}'-\mathbf{s}'\right|^{2}}\big\{ \bnabla_{\left(\bf{s},\bf{s}'\right)}\cdot\left[\braket{\hat{\rho}\!\left({\bf s}\right)\hat{\rho}\!\left({\bf s}'\right)}_{c}\bnabla_{\left(\bf{s},\bf{s}'\right)}\left[V\!\left({\bf s}\right)+V\!\left({\bf s}'\right)\right]\right]\big.\nonumber \\
 & -l_\text{r}\bnabla_{\left(\bf{s},\bf{s}'\right)}\bnabla_{\left(\bf{s},\bf{s}'\right)}:\left\{\left[\bnabla_{{\bf s}}V\!\left({\bf s}\right)\right]\braket{\hat{\rho}\!\left({\bf s}'\right)\hat{\mb}\left({\bf s}\right)}_{c}\oplus\left[\bnabla_{{\bf s}'}V\!\left({\bf s}'\right)\right]\braket{\hat{\rho}\!\left({\bf s}\right)\hat{\mb}\left({\bf s}'\right)}_{c}\right\}\nonumber \\
 & \big.+T\delta\!\left(\mathbf{s}-\mathbf{s}'\right)\nabla_{{\bf s}}^{2}\rho\!\left({\bf s}\right)+l_\text{r}T\bnabla_{\left(\bf{s},\bf{s}'\right)}\cdot\delta\!\left({\bf s}-{\bf s}'\right)\nabla^2_{\left({\bf s},{\bf s}'\right)}\left[\mb\!\left({\bf s}'\right)\oplus\mb\!\left({\bf s}\right)\right]\big\} \nonumber \\
 & +\mo{U_{0}\rho_{b}^{3},r^{-4}},\label{eq:P2cm}
\end{align}
where we have utilized the homogeneous phase boundary condition, which
gives $\lim_{r,r'\rightarrow\infty}P_{c}^{\left(2\right)}\!\left(\br,\br'\right)=0$.
By the above considerations, $P_{c}^{\left(2\right)}$ has an asymptotic behavior of a localized dipole, {\it i.e.} $P_{c}^{\left(2\right)}\sim \min\!\left(r^{-3},r'^{-3}\right)$. From this point, one can invert the expansion to obtain $\braket{\hat{\rho}\!\left(\br\right)\hat{\rho}\!\left(\br'\right)}_{c}\sim\min\!\left(r^{-3},r'^{-3}\right)$,
as done for $\rho\!\left(\br\right)$ in Sec.~\ref{sec:far-field}. Following this procedure would require to assume a stress expansion of the form
\begin{align}
\bm{\sigma}^{\left(2\right)}\!\left(\br,\br'\right)=\bm{\sigma}^{\left(2\right)}\!\left(\braket{\hat{\rho}\!\left(\br\right)\hat{\rho}\!\left(\br'\right)}\right)+\mo{\left(\partial+\partial'\right)\braket{\hat{\rho}\!\left(\br\right)\hat{\rho}\!\left(\br'\right)}},
\end{align}
which can be proved by computing the dynamics of three-point correlations, {\it e.g.} $\braket{\hat{\rho}\!\left(\br\right)\hat{\rho}\!\left(\br'\right)\hat{\rho}\!\left(\br''\right)}$, and truncating the expansion at the next order by omitting four-point correlations.

\subsection{Derivation of the virial expansion} \label{sec:derv-vir}

We now show that
the second virial coefficient in Eq.~\eqref{eq:vdw} is $1/2$.
To this end, we need to calculate the first-order correction to pressure
due to the leading-order behaviors of the stress components
$\bm\sigma^\text{IK}$ and $\bm\sigma^\text{P}$ originating from
the interactions between particles.

We first calculate $\bm{\sigma}^\text{IK}$ up to the leading order.
In the weak-interaction regime, as previously discussed,
$\braket{\hat{\rho}\!\left(\br\right)\hat{\rho}\!\left(\br'\right)}=\rho\!\left(\br\right)\rho\!\left(\br'\right)+\mathcal{O}\!\left(\min\!\left(r^{-3},r'^{-3}\right),U_{0}\rho_{b}^{3}\right)$ holds in the far field.
Applying this approximation, Eq.~\eqref{eq:IK} can be expanded as
\begin{align}
\bm{\sigma}^\text{IK}\!\left(\br\right) & =\frac{1}{2}\int d^{2}\br'\frac{\br'\br'}{r'}\,\frac{dU\!\left(r'\right)}{dr'}\int_{0}^{1}d\lambda\,\rho\!\left(\br+\left(1-\lambda\right)\br'\right)\rho\!\left(\br-\lambda\br'\right)+\mathcal{O}\!\left(r^{-3},U_{0}^{2}\rho_{b}^{3}\right).\label{eq:IK-1}
\end{align}
To simplify this expression further, we note that the integral over $\lambda$ contains densities which can be expanded as
\begin{align}
\rho\!\left(\br+\left(1-\lambda\right)\br'\right) & =\rho\!\left(\br\right)+\left(1-\lambda\right)\br'\cdot\bnabla\rho\!\left(\br\right)+\mathcal{O}\!\left[\left(1-\lambda\right)^{2}r'^2\partial^{2}\rho\right]
\end{align}
and
\begin{align}
\rho\!\left(\br-\lambda\br'\right) & =\rho\!\left(\br\right)-\lambda\br'\cdot\bnabla\rho\!\left(\br\right)+\mathcal{O}\!\left(\lambda^{2}r'^2\partial^{2}\rho\right).
\end{align}
Substituting these expansions into Eq.~\eqref{eq:IK-1}
and carrying out the integration over $\lambda$, we obtain
\begin{align}
\bm{\sigma}^\text{IK}\!\left(\br\right) & =\frac{\rho\!\left(\br\right)^2}{2}\int d^{2}\br'\frac{\br'\br'}{r'}\,\frac{dU\!\left(r'\right)}{dr'}+\mathcal{O}\!\left(r^{-3},U_{0}^{2}\rho_{b}^{3}\right),\label{eq:IKmf}
\end{align}
where we have used the far-field behavior
$\partial^{2}\rho\sim r^{-3}$ derived from Eq.~\eqref{eq:ffrho}.
After evaluating the area integral over $\br'$ using integration by parts,
we find
\begin{align}
\bm{\sigma}^\text{IK} & =-\frac{U_{0}}{2}\rho^{2}\id+\mo{r^{-3},U_{0}^{2}\rho_{b}^{3}}\nonumber \\
 & =-\frac{U_{0}}{2}\rho_{b}^{2}\id-U_{0}\rho_{b}\left(\rho-\rho_{b}\right)\id+\mo{r^{-3},U_{0}^{2}\rho_{b}^{3}}.\label{eq:ik-1}
\end{align}
Thus the contribution of $\bm\sigma^\text{IK}$ to the bulk pressure,
or the direct interaction pressure $P_\text{D}\!\left(\rho_{b}\right)=-\tr\bm{\sigma}^\text{IK}\!\left(\rho_{b}\right)/2$,
satisfies
\begin{align}
\frac{P_\text{D}\!\left(\rho_{b}\right)}{T_\text{eff}\,\rho_{b}} & =\frac{U_{0}\rho_{b}}{2T_\text{eff}}+\mathcal{O}\!\left[\left(\frac{U_{0}\rho_{b}}{T_\text{eff}}\right)^{2}\right],\label{eq:Pd1}
\end{align}
which is an exact analog of the leading-order contribution of
interparticle interactions to the bulk pressure in a passive gas,
the only change being the replacement of temperature with $T_\text{eff}$.

We now turn to the leading-order behavior of $\bm{\sigma}^\text{P}$,
which can be obtained similarly as follows.
Again assuming the weak-interaction regime,
we can use the previously obtained relation
$\braket{\hat{\rho}\!\left(\br\right)\hat{\mb}^{\left(n\right)}\left(\br'\right)}=\rho\!\left(\br\right)\mb^{\left(n\right)}\!\left(\br'\right)+\mo{r^{-3},r'^{-3},U_{0}\rho_{b}^{3}}$,
so that Eq.~\eqref{eq:sigmap} can be expanded as
\begin{align}
\bm{\sigma}^\text{P}\!\left(\br\right) & =l_\text{r}\int d^{2}\mathbf{r}'\left[\bnabla U\!\left(\left|\mathbf{r}-\mathbf{r}'\right|\right)\right]\rho\!\left(\mathbf{r}'\right)\mb\!\left(\br\right)+Tl_\text{r}\bm{\nabla}\mathbf{m}-2\left(T_\text{eff}-T\right)\mathbb{Q}+\mo{r^{-3},U_{0}^{2}\rho_{b}^{3}}.\label{eq:sigmapmf}
\end{align}
Using integrating by parts,
the area integral over $\br'$ can be rewritten as
\begin{align}
\int d^{2}\mathbf{r}'\,\rho\!\left(\mathbf{r}'\right)\bnabla U\!\left(\left|\mathbf{r}-\mathbf{r}'\right|\right) &= -\int d^{2}\mathbf{r}'\,U\!\left(\left|\mathbf{r}-\mathbf{r}'\right|\right)\bnabla'\rho\!\left(\mathbf{r}'\right).
\end{align}
Using this relation in Eq.~\eqref{eq:sigmapmf}
and expanding $\rho\!\left(\br'\right)$ about $\br'=\br$, we can evaluate
the area integral over $\br'$ to obtain
\begin{align}
\bm{\sigma}^\text{P} & =l_\text{r}U_{0}\left(\bnabla\rho\right)\mb+Tl_\text{r}\bm{\nabla}\mathbf{m}-2\left(T_\text{eff}-T\right)\mathbb{Q}+\mo{r^{-3},U_{0}^{2}\rho_{b}^{3}},
\end{align}
where we again used the far-field behavior $\partial^{2}\rho\sim r^{-3}$.
This implies that, at order $U_0 \rho_b$,
$\bm{\sigma}^\text{P}$ vanishes in the bulk. As a result,
the contribution of $\bm\sigma^\text{P}$ to the bulk pressure,
or the indirect interaction pressure, satisfies
$P_\text{I}\!\left(\rho_{b}\right)/(\rho_b T_\text{eff})=\mo{U_{0}^{2}\rho_{b}^{2}/T_\text{eff}^{2}}$.
Using this result together with Eqs.~\eqref{eq:Pdec} and~\eqref{eq:Pd1},
we finally obtain the virial expansion~\eqref{eq:vdw}.
The derivation we have presented so far clearly shows that
the virial expansions for both active and passive particles
coincide up to the first order
(only with the usual temperature replaced by an effective temperature)
because the indirect pressure $P_\text{I}$,
which captures the effects of ``swimming'',
only contributes higher-order corrections.

\section{Scalar, vector and tensor shear stresses\label{sec:Scalar-and-vector}}

As stated in the main text, the long-distance decay of the traceless deviatoric stress tensor $\mathbb{S}$ satisfies
\begin{align}
\mathbb{S} & =\mathcal{O}\!\left\{ \left[P-P\!\left(\rho_{b}\right)\right]^{2},\partial P\right\} ,\label{eq:Sdecay}
\end{align}
and $\bnabla\cdot\mathbb{S}$ admits the Helmholtz decomposition
\begin{align}
\bnabla\cdot\mathbb{S} & =-\bnabla\Phi_{S}+\bnabla\times\bm{\Psi},\nonumber
\end{align}
stated in Eq.~\eqref{eq:helm-1}.
Here we show that both $\Phi_S$ and $\bm\Psi$ decay with the distance as $\mathcal{O}\!\left(\mathbb{S},r^{-2}\right)$, justifying Eq.~\eqref{eq:phiP}.
This is not a trivial statement---due to the nonlocal nature
of the Helmholtz decomposition for vectors, Eq.~\eqref{eq:helm-1}
does not immediately guarantee that $\mathbb{S}$, $\Phi_S$,
and $\bm\Psi$ are of the same order. In the following, we address
this difficulty by applying a tensor version of
the Helmholtz decomposition.

As the first step, we decompose $\mathbb{S}$ as
\begin{align} \label{eq:S_decom}
\mathbb{S} & =\mathbb{A}+\mathbb{E},
\end{align}
where $\mathbb{A}\equiv\left(\mathbb{S}-\mathbb{S}^{T}\right)/2$ and 
$\mathbb{E}\equiv\left(\mathbb{S}+\mathbb{S}^{T}\right)/2$ are
the antisymmetric and the symmetric components of $\mathbb{S}$,
respectively. In analogy to linear flow, $\mathbb{A}$
can be thought of as a pure rotation, while $\mathbb{E}$ as a pure straining
motion\cite{BergenholtzJFM2002}. We note that, among the components of
the stress tensor $\bm\sigma$ shown in Eq.~\eqref{eq:sigma},
only the polarization component $\bm{\sigma}^\text{P}$
is not symmetric and can thus contribute to $\mathbb{A}$,
see Eq.~\eqref{eq:sigmap}. Because both $\mathbb{A}$ and $\mathbb{E}$ are local functions of $\mathbb{S}$, it is evident that $\mathbb{A}=\mo{\mathbb{S}}$ and $\mathbb{E}=\mo{\mathbb{S}}$. It remains to show that this decay is inherited by their contributions to $\Phi_S$ and $\bm{\Psi}$. We will first show this for the contributions by $\mathbb{A}$, and then for the contributions by $\mathbb{E}$.

Due to the constraint of antisymmetry,
the rank-$2$ tensor $\mathbb{A}$ has only a single free parameter,
which allows the following representation:
\begin{align} \label{eq:A_comps}
A_{\alpha\beta} & =-\varepsilon_{\alpha\beta\gamma}\Omega_{\gamma},
\end{align}
where $\bm{\Omega}=\Omega\mathbf{e}_{z}$.
Then we can write $\bnabla\cdot\mathbb{A}=\bnabla\times\bm{\Omega}$,
which means that $\bnabla\cdot\mathbb{A}$ contributes only to
the solenoidal component $\bnabla\times\bm{\Psi}$ of
$\bnabla\cdot\mathbb{S}$. Moreover, since the above representation
can be inverted as $\Omega_\alpha = \varepsilon_{\alpha\beta\gamma} A_{\beta\gamma}/2$,
$\bm{\Omega}$ is clearly a local linear function of $\mathbb{S}$.
Thus, Eq.~\eqref{eq:Sdecay} implies
\begin{align} \label{eq:A_far-field}
\bm{\Omega} = \mathcal{O}(\mathbb{A}) = \mathcal{O}(\mathbb{S}) = \mathcal{O}\!\left\{ \left[P-P\!\left(\rho_{b}\right)\right]^{2},\partial P\right\}
\end{align}
Hence, $\mathbb{A}$ cannot contribute to $\Phi$ defined in Eq.~\eqref{eq:scalars}, and its contributions are bound to
be higher-order than the leading-order terms of Eq.~\eqref{eq:phiP}.

Now, it remains to show that the contributions by $\mathbb{E}$ also decay with distance
in the same way. Applying a tensor version of the Helmholtz 
decomposition, also called the generalized Beltrami decomposition~\cite{Beltrami1892,Gurtin1963,Fosdick2003,Fosdick2005,Admal2016},
the symmetric component $\mathbb{E}$ can be decomposed as
\begin{align}
\mathbb{E} & =\mathbb{E}^\text{S}+\mathbb{E}^\text{I},\label{eq:belt}
\end{align}
where
\begin{align}
\mathbb{E}^\text{S} & =\bnabla\times\left(\bnabla\times\bm{\Pi}\right),\label{eq:Es-2}\\
\mathbb{E}^\text{I} & =\frac{1}{2}\left[\bnabla{\bf v}+(\bnabla{\bf v})^{T}\right]\label{eq:Es-1-1}
\end{align}
for a symmetric rank-$2$ tensor $\Pi$ and a vector potential $\bf v$.
These imply $\bnabla\cdot\mathbb{E}^{S}=0$ and $\bnabla\times\left(\bnabla\times\mathbb{E}^{I}\right)=0$.
Conversely, $\bnabla\cdot\mathbb{E}=0$ implies $\mathbb{E}=\mathbb{E}^\text{S}$,
and $\bnabla\times\left(\bnabla\times\mathbb{E}\right)=0$ implies
$\mathbb{E}=\mathbb{E}^\text{I}$. Thus, $\mathbb{E}^\text{S}$ can be
regarded as the solenoidal component of $\mathbb{E}$,
and $\mathbb{E}^\text{I}$ the irrotational component. For example, linear fluids correspond to the case $\mathbb{E}=\mathbb{E}^{I}$,
where ${\bf v}={\bf J}/\rho$ is the fluid velocity.

Recently, it has been shown that the generalized Beltrami decomposition
satisfies the following integrability rule\cite{Maggiani2015}:
defining $\abs{\mathbb{E}} \equiv \sqrt{E_{\alpha\beta}E_{\alpha\beta}}$, if $\int d^{2}\br\abs{\mathbb{E}}^{p}<\infty$ for some fixed $p>1$,
then we also have $\int d^{2}\br\abs{\mathbb{E}^\text{S}}^{p}<\infty$ and $\int d^{2}\br\abs{\mathbb{E}^\text{I}}^{p}<\infty$. This stems from the fact that the space of symmteric tensors $\mathbb{E}$ satisfying $\int d^{2}\br\abs{\mathbb{E}}^{p}<\infty$ can be decomposed into a direct sum of two subspaces -- one being the subspace of all irrotational tensors, and the other being the subspace of all solenoidal tensors. For the special case $p=2$, this can be seen immediately, as the decomposition becomes an orthogonal one. Using integration by parts, one can verify that a tensor orthogonal to $\mathbb{E}^S$ defined in Eq.~\eqref{eq:Es-2} is of the form Eq.~\eqref{eq:Es-1-1}, with the orthogonality taken under the standard inner product $\braket{\mathbb{E}^1|\mathbb{E}^2}\equiv\int d^2 \br \thinspace E^1_{\alpha\beta}\left(\br\right)E^2_{\alpha\beta}\left(\br\right)$~\cite{Admal2016}. Note that, for two-dimensional smooth fields $\mathbb{E}$ whose derivatives
vanish as $r\rightarrow\infty$,
$\int d^{2}\br\abs{\mathbb{E}}^{p}<\infty$ holds
if and only if $\abs{\mathbb{E}}^{p}=o\!\left(r^{-2}\right)$.
This is equivalent to the requirement $\mathbb{E}=o\!\left(r^{-2/p}\right)$. Using the notation $\gamma=2/p$, we can rewrite
the integrability rule as follows:
\begin{theorem*}
If $\mathbb{E}=o\!\left(r^{-\gamma}\right)$
for some fixed $0<\gamma<2$, then $\mathbb{E}^\mathrm{S}=o\!\left(r^{-\gamma}\right)$ and $\mathbb{E}^\mathrm{I}=o\!\left(r^{-\gamma}\right)$ also hold.
\end{theorem*}

This result can be refined further for our purpose. To proceed, we suppose $\mathbb{E}=\mathcal{O}\left(r^{-\gamma}\right)$ for some
$\gamma>0$. Note that we expect $P-P\!\left(\rho_{b}\right)=\mo{r^{-1}}$,
which would correspond, according to Eq.~\eqref{eq:Sdecay}, $\gamma=2$. Indeed, we will show using the general exponent $\gamma$ that this is the case. First, we denote $\gamma_\text{cf}\equiv\min{\left(\gamma,2\right)}$. Then, it holds that $\mathbb{E}=\mathcal{O}\left(r^{-\gamma_\text{cf}}\right)$. In particular, for any $0<\gamma'<\gamma_\text{cf}$, it is true that $\mathbb{E}=o(r^{-\gamma'})$. By our integrability rule, $\mathbb{E}^{S}=o\left(r^{-\gamma'}\right)$
and $\mathbb{E}^{I}=o\left(r^{-\gamma'}\right)$. Taking the limit $\gamma'\rightarrow\gamma_\text{cf}$, we obtain  $\mathbb{E}^{S}=\mathcal{O}\left(r^{-\gamma_\text{cf}}\right)$
and $\mathbb{E}^{I}=\mathcal{O}\left(r^{-\gamma_\text{cf}}\right)$, up to some sub-algebraic modulation of the decay.

Put differently, we have found that $\mathbb{E}^\text{I}$ is of order $\mathcal{O}(\mathbb{E},r^{-2})$. To apply this result to the far-field behaviors of $\Phi_S$ and $\bm\Psi$, we go back to Eq.~\eqref{eq:helm-1} and examine the far-field behavior of $\bnabla\cdot\mathbb{S}$, which is dominated by $\bnabla\cdot\mathbb{E}$, as already discussed. Taking the divergence of Eq.~\eqref{eq:belt} side by side, the solenoidal component $\mathbb{E}^\text{S}$ vanishes, leaving
\begin{align}
\bnabla\cdot\mathbb{E} & =\bnabla\left(\bnabla\cdot{\bf v}\right)+\frac{1}{2}\left[\lap{\bf v}-\bnabla\left(\bnabla\cdot{\bf v}\right)\right].\label{eq:beltd}
\end{align}
On the rhs, one can easily find that $\bnabla\left(\bnabla\cdot{\bf v}\right)$ is the irrotational component, while
\begin{align}
\frac{1}{2}\left[\lap{\bf v}-\bnabla\left(\bnabla\cdot{\bf v}\right)\right] = -\frac{1}{2}\bnabla\times(\bnabla\times \bv)
\end{align}
is the solenoidal component. Combining these observations with Eqs.~\eqref{eq:helm-1}, \eqref{eq:S_decom}, \eqref{eq:A_comps}, and \eqref{eq:belt}, we identify
\begin{align}
\Phi_S = -\bnabla\cdot \bv, \quad \bm\Psi = \bm\Omega - \frac{1}{2}\bnabla\times\bv.	
\end{align}
From Eq.~\eqref{eq:Es-1-1} and the far-field behavior of $\mathbb{E}^\text{I}$, we obtain $\bnabla\cdot{\bf v}=\tr\,\mathbb{E}^\text{I}=\mathcal{O}\!\left(\mathbb{E},r^{-2}\right) = \mo{\mathbb{S},r^{-2}}$ up to a sub-algebraic modulation.
Then, using the above identities, we finally conclude that $\Phi_{S}=\mathcal{O}(\mathbb{S},r^{-2})$ and $\bm\Psi=\mathcal{O}(\mathbb{S},r^{-2})$. We have thus confirmed Eq.~\eqref{eq:phiP}.

\section{Finite-size effects\label{sec:Finite-Size-Effects}}

Here we address two different issues about how the infinite-size limit is
achieved. First, we clarify the meaning of the infinite-area integral
appearing in the current-force relation~\eqref{eq:cfr}.
Second, we briefly discuss how the finite-size effects modify the derivations
shown in Sec.~\ref{sec:far-field}, which are valid in the infinite-size limit.
As an explicit example, we show that the dipole moment of a single
asymmetric body in an $L\times L$ torus converges algebraically to the
asymptotic value as $L\to\infty$.

\subsection{Derivation of the current-force relation~\eqref{eq:cfr}}

Integrating Eq.~\eqref{eq:Jff} side by side over the entire space, we obtain
\begin{align} \label{eq:cfrfin}
\int d^{2}\br\,\bj\!\left(\br\right) & =\frac{\mu}{2}\bp,
\end{align}
which differs by a factor of $1/2$ from the well-established current-force
relation~\eqref{eq:cfr}. As discussed below, this apparent contradiction
is resolved if one properly defines the area integral over the entire system
appearing in Eq.~\eqref{eq:cfr}.

By integrating Eq.~\eqref{eq:cdyn-1} side by side over an area
$\mathcal{A}$ which contains all the bodies inside, the divergence theorem
and $V = 0$ on the boundary imply 
\begin{align}
\int_{\mathcal{A}} d^{2}\br\,\mathbf{J}\!\left(\br\right) & =\mu\bp+\mu\oint_{\partial\mathcal{A}} d\ell\,\mathbf{e}_{n}\cdot\bm{\sigma}\!\left(\br\right),\label{eq:cint}
\end{align}
where $d\ell$ is an infinitesimal segment on the boundary
$\partial\mathcal{A}$, and $\mathbf{e}_{n}$ is a unit normal vector.
For a finite system with periodic boundaries,
if $\mathcal{A}$ covers the entire system,
the boundary integral in Eq.~\eqref{eq:cint} is carried out twice
for each $d\ell$ with opposite directions of $\mathbf{e}_{n}$,
so that its value sums to zero. As long as $\mathcal{A}$ covers the entire
system, the same result still holds even in the limit $L\to\infty$.
The infinite-area integral in Eq.~\eqref{eq:cfr} should be interpreted
in this vein---the infinite-size limit is taken after requiring that
$\mathcal{A}$ covers the entire system.

How do we then obtain Eq.~\eqref{eq:cfrfin} as well?
Going back to Eq.~\eqref{eq:cint}, we choose $\mathcal{A}$ to be a disk
$\mathcal{D}_{R}$ of radius $R$ centered at the origin,
take the infinite-size limit, after which $R$ is sent to infinity.
Using this order of limits, we can write
\begin{align}
\mu\oint_{\partial\mathcal{D}_{R}} d\ell\,\mathbf{e}_{n}\cdot\bm{\sigma}\!\left(\br\right) = \mu\int_{\mathcal{D}_{R}} d^2\br\,\bm\nabla\cdot \bm\sigma(\mathbf{r}) = -\frac{\mu}{2}\mathbf{p} + \mathcal{O}\!\left(R^{-1}\right),
\end{align}
where the last equality is obtained by using Eq.~\eqref{eq:sigexp-1}
to evaluate $\bm\nabla\cdot\bm\sigma(\br)$.
Using this relation in Eq.~\eqref{eq:cint}, we obtain
\begin{align}
\int_{\mathcal{D}_{R}}d^{2}r\,\mathbf{J}\!\left(\br\right) & =\frac{\mu}{2}\bp+\mathcal{O}\!\left(R^{-1}\right),\label{eq:pcint}
\end{align}
which gives the precise meaning of Eq.~\eqref{eq:cfrfin}. To sum up,
whether one gets Eq.~\eqref{eq:cfr} or Eq.~\eqref{eq:cfrfin} is determined
by whether the area integral expands with or slower than the system size.

\subsection{Finite-size corrections in a periodic system}

For a finite system $\mathcal{S}$, the proper solution for
Eqs.~\eqref{eq:lapphi} and \eqref{eq:pois} is not Eq.~\eqref{eq:green},
but (see, for example, Ref.~\cite{Griffiths1962})
\begin{align}
\Phi\!\left(\br\right)&=P\!\left(\rho_{b}\right)-\frac{1}{2\pi}\int_{\mathcal{S}}d^{2}\mathbf{r}'\ln\left|\mathbf{r}-\mathbf{r}'\right|\left\{ \bnabla'\cdot\left[\rho\!\left(\br'\right)\bnabla'V\!\left(\br'\right)\right]-l_\text{r}\partial_{\alpha}'\partial_{\beta}'\left[m_{\alpha}\!\left(\br'\right)\partial_{\beta}'V\!\left(\br'\right)\right]\right\} \nonumber \\
 &\quad +\frac{1}{2\pi}\oint_{\partial\mathcal{S}} d\ell\,\ln\abs{\br-\br_\ell'}\left(\mathbf{e}_{n}\right)_{\alpha}\partial_{\beta}'\sigma_{\alpha\beta}\!\left(\br_\ell'\right),\label{eq:greenfin}
\end{align}
where $\br_\ell'$ in the second integral is on the boundary segment $d\ell$.
The boundary integral on the second line is indeed responsible for the
finite-size effects observed in Fig.~\ref{fig:numerics-dens-current} near
the boundary. Since it would be physically absurd if the stress diverges
with the distance from the origin, it is reasonable to require that
$\bnabla\cdot\bm{\sigma}(\mathbf{r}) = o\!\left(r^{-1}\right)$.
This implies that the boundary integral is $o\!\left(1\right)$,
so that the derivations in Sec.~\ref{sec:far-field} are fully valid
in the infinite-size limit.

Precisely how the boundary contributions decay with the increasing system size
could be dependent on the details of the system and its boundary conditions.
As an explicit example, below we show for the dipole moment that these corrections do decay with
the system size $L$, namely $\mathcal{O}\!\left(L^{-2}\right)$.

\begin{figure}
  \centering
  \includegraphics[width=0.4\textwidth]{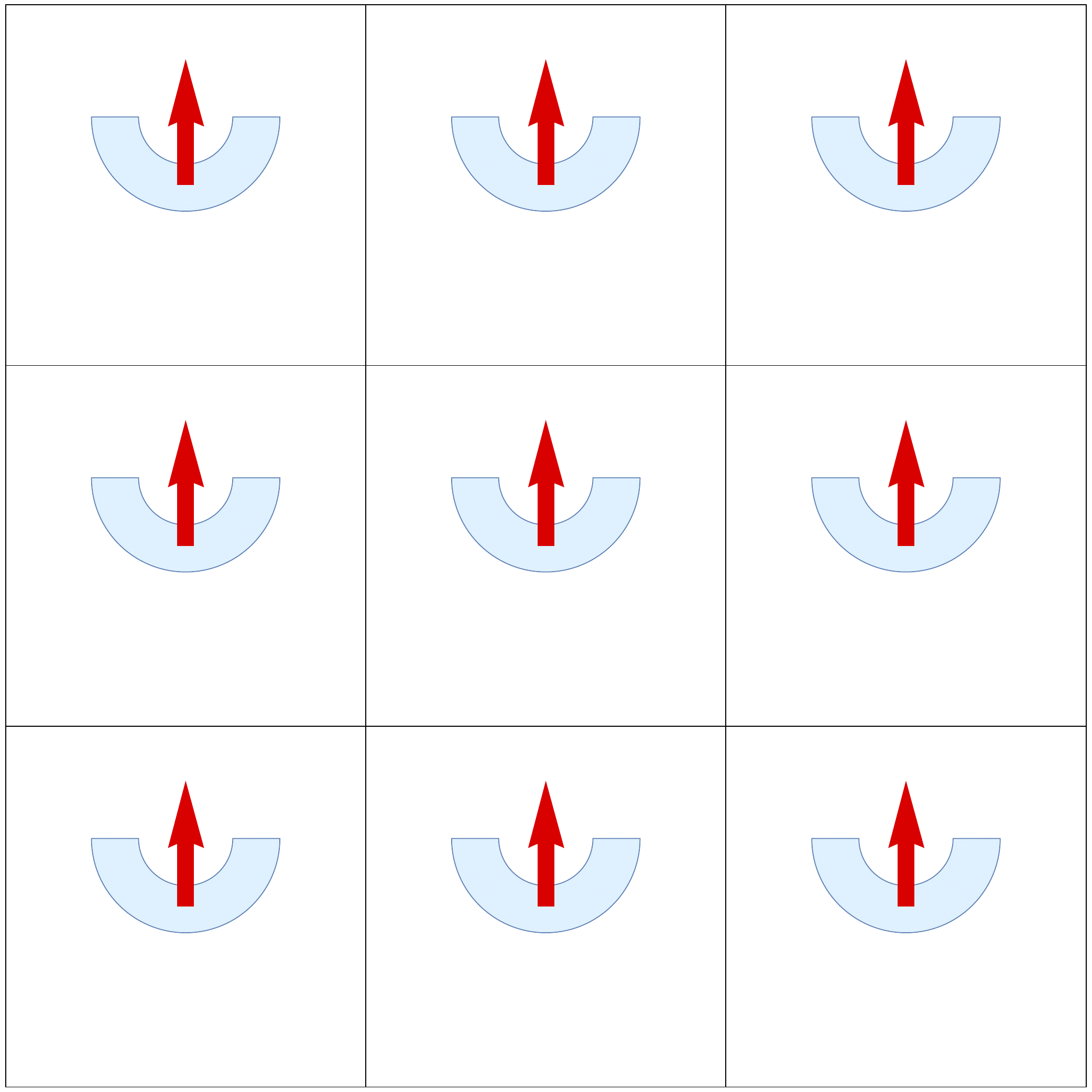}
  \caption{A section from a cubic lattice of identical passive bodies. Dipole moments are depicted in red.}
  \label{fig:lattice}
\end{figure}

We consider a single body described by a potential $V$ in a periodic torus
of dimensions $L\times L$. Extension to mutually distant multiple bodies
is straightforward. Furthermore, we assume that the boundaries are
in the far field of the body, so that finite-size effects can be described
using far-field effects. Given these assumptions,
the system can be regarded as an infinite cubic
lattice with lattice constant $L$, where an exact copy of the body is placed
at the center of each cell (see Fig.~\ref{fig:lattice}). The lattice is now characterized by a periodic
potential $V=\sum_{i}V_{i}$. We denote by $\bp_{i}\equiv-\int d^{2}\mathbf{r}\rho\bnabla V_{i}$
the force applied to the fluid by body copy $i$, and $\mathbf{R}_{i}$
represents the corresponding response tensor. We denote by $\bp$
and ${\bf R}$ as the same quantities in the $L\to\infty$ limit, respectively.
Following the procedure described in Sec.~\ref{sec:interactions}, we obtain
\begin{align}
\bp_{j} & =\bp+\mathbf{R}_{j}\sum_{i\neq j}\frac{1}{2\pi}\frac{\mathbf{r}_{ij}\cdot\bp_{i}}{r_{ij}^{2}}+\sum_{i\neq j}\mathcal{O}\!\left(r_{ij}^{-2}\right).
\end{align}
Since the lattice constant is $L$, $r_{ij}\sim n_{ij}L$
with $n_{ij}$ designating the rescaled distance between body $i$ and $j$.
Thus we have $\bp_{i}=\bp+\mathcal{O}\!\left(L^{-1}\right)$ for any $i$ and $\sum_{i}\mathcal{O}\!\left(r_{ij}^{-2}\right)=\mathcal{O}\!\left(L^{-2}\right)$. Noting that the potentials $V_i$ are all identical to each other, Eq.~\eqref{eq:Rjrho} implies that the single-body response coefficient $\mathbf{R}_i=\mathbf{R}$ for all $i$. Thus, the above equation can be rewritten as
\begin{align}
\bp_{j} & =\bp+\mathbf{R}\sum_{i\neq j}\frac{1}{2\pi}\frac{\mathbf{r}_{ij}\cdot\bp}{r_{ij}^{2}}+\mathcal{O}\!\left(L^{-2}\right).
\end{align}
By reflection symmetry, the first-order terms should vanish;
thus we finally obtain
\begin{align}
\bp_{j} & =\bp+\mathcal{O}\!\left(L^{-2}\right).\label{eq:prolast}
\end{align}

\bibliographystyle{jstat}
\begingroup
\raggedright

\bibliography{manuscript}
\endgroup

\end{document}